\documentclass[useAMS,usenatbib]{mn2e}
\usepackage{amssymb, amsmath, epsfig}
\newcommand{\bfk}{{\mathbf{k}}}

\newcommand{\barxi}{{\bar{x}_i}}

\newcommand{\bfC}{{\boldsymbol{C}}}

\newcommand{\bfx}{{\bf x}}

\newcommand{\nhat}{{\hat{\mathbf{n}}}}
\newcommand{\Mpc}{{\rm Mpc}}

\newcommand{\Msun}{{M_{\odot}}}
\def\apj{ApJ}
\def\apjl{ApJL}
\def\apjs{ApJS}
\def\aj{AJ}
\def\mnras{MNRAS}
\def\pasj{PASJ}

\def\aap{{\em A.\&A}}

\begin{document}

\title{Studying Reionization with Ly$\alpha$ Emitters}

\author[M. McQuinn et al.]{Matthew McQuinn$^1$\thanks{mmcquinn@cfa.harvard.edu},
Lars Hernquist$^{1}$, 
Matias Zaldarriaga$^{1,2}$, 
Suvendra Dutta$^{1}$\\
$^{1}$ Harvard-Smithsonian Center
for Astrophysics, 60 Garden St., Cambridge, MA 02138\\ 
$^2$  Jefferson Laboratory of Physics, Harvard University,
Cambridge, MA 02138\\ 
}

\pubyear{2006} \volume{000} \pagerange{1}

\maketitle\label{firstpage}

\begin{abstract}

We show that observations of high-redshift Ly$\alpha$ emitters (LAEs)
have the potential to provide definitive evidence for reionization in
the near future.  Using $200$ Mpc radiative transfer simulations, we
calculate the effect that patchy reionization has on the line profile,
on the luminosity function, and, most interestingly, on the clustering
of emitters for several realistic models of reionization.
Reionization increases the measured clustering of emitters, and we
show that this enhancement would be essentially impossible to
attribute to anything other than reionization.  Our results motivate
looking for the signature of reionization in existing LAE data.  We
find that for stellar reionization scenarios the angular correlation
function of the $58$ LAEs in the Subaru Deep Field $z = 6.6$
photometric sample is more consistent with a fully ionized universe
(mean volume ionized fraction $\bar{x}_i \approx 1$) than a universe
with $\bar{x}_i < 0.5$ at $>2$-$\sigma$ confidence level.
Measurements in the next year on Subaru will increase their $z =
6.6$ LAE sample by a factor of five and tighten these limits.  If the
clustering signature of reionization is detected in a LAE survey, a
comparison with a Lyman-break or a H$\alpha$ survey in the same field
would confirm the reionization hypothesis.  We discuss the optimal LAE
survey specifications for detecting reionization, with reference to
upcoming programs.

\end{abstract}


\section{Introduction}
The reionization epoch -- when the hydrogen in the Universe was
ionized by photons produced in the first galaxies -- remains one of
the least explored periods in cosmology.  There is substantial
uncertainty as to when reionization occurred.  The latest cosmic
microwave background (CMB) data show that the mean redshift of
reionization was $z_{\rm rei} = 11 \pm 3$, but are consistent with
$z_{\rm rei} = 0$ at 3-$\sigma$ \citep{page06}, and measurements of
Ly$\alpha$ absorption in the spectra of high-redshift quasars suggest
that reionization ended at $z \approx 6$ \citep{becker01, white03,
fan06}.

Whether observations have already pinned down the redshift of bubble
percolation to be $z \approx 6$ using the Ly$\alpha$ absorption
features in the spectra of high-redshift quasars -- the Ly$\alpha$
forest -- is under debate.  The appearance of Gunn-Peterson
absorption troughs in the $z > 6$ Ly$\alpha$ forest may
signify a change in the ionization state of the intergalactic medium
\citep{becker01, white03, fan06}.  In addition, the rapid decrease of
Ly$\alpha$ forest transmission at $z > 6$ (as well as the decrease in
transmission in the Ly$\beta$ and the Ly$\gamma$ forests), the large
variance in the transmission between sight-lines, and the sizes of the
proximity regions around $z\approx 6$ QSOs may also indicate that
reionization is ending (e.g., \citealt{becker01, white03,fan06,
wyithe05, wyithe05b, mesinger04, mesinger07}). 

However, owing to the resonant nature of the Ly$\alpha$ line,
Ly$\alpha$ transmission is essentially zero if the gas is more neutral
than one part in a thousand, which makes the Ly$\alpha$ forest
insensitive to the order unity fluctuations in the ionized fraction
that define reionization.  Because of this limitation, arguments for
the end of reionization based on the forest are
controversial. \citet{becker06} proposed that the rapid evolution of
Ly$\alpha$ transmission in the $z \approx 6$ forest is consistent with an
ionized intergalactic medium (IGM).  \citet{lidz06a} and \citet{lui06}
showed that the observed variance in the transmission between
sight-lines is consistent with the variance one expects from an
ionized IGM, and \citet{bolton06} and \citet{lidz07} demonstrated that
current data cannot distinguish the proximity region owing to enhanced
ionizing flux from the effect of an HII region.

High-redshift QSOs will probably not reveal additional information
about reionization in the next few years, as all-sky surveys that
observe in the near-infrared are required to push to higher redshifts.
The Ly$\alpha$ forest in the spectrum of high-redshift gamma ray
bursts (GRBs) holds more immediate promise, but the presence of damped
Ly$\alpha$ systems within GRB host galaxies complicates the
determination of the ionization state of the IGM.  Despite this
difficulty, through measurements of the Ly$\alpha$ and the Ly$\beta$
forests in the spectrum of a single GRB, \citet{totani05} derived the
constraint $\bar{x}_i > 0.4$ at $z = 6.3$, assuming a uniformly
ionized IGM, where $\bar{x}_i$ is the global ionized fraction of the
IGM.  Accounting for a patchy reionization process will weaken this
limit.

Observations of high-redshift 21cm emission have the potential to put
the strongest constraints on $\bar{x}_i$ in the long term.  In
principle, such studies can image the neutral hydrogen in the Universe
as it became ionized.  However, 21cm observations must first overcome
terrestrial radio broadcasts as well as galactic foregrounds that are
four orders of magnitude larger than the signal (e.g.,
\citealt{zaldarriaga04, mcquinn06, furl-rev}).  Despite these
difficulties, the Mileura Widefield Array and the Low Frequency Array
will start dedicated observations of this emission in the coming years
\citep{morales03, deVos04}, and the 21cm Array and the Giant Meter-wave
Radio Telescope are already being used to perform these measurements
\citep{pen04}.

Studies of the CMB will never be able to provide a direct detection of
reionization by imaging individual HII regions. However, measurements
of CMB anisotropies can improve constraints on the global properties
of reionization.  A more precise measurement of CMB polarization
anisotropies at low multipoles has the potential to provide some
information about the duration of reionization in addition to tighter
constraints on the average redshift of reionization (e.g.,
\citealt{keating05}). The {\it Planck} mission will measure the
large-scale E-mode polarization anisotropies to nearly the
cosmic-variance limit in the next few years. In addition, a sizable
fraction of arc-minute-scale CMB anisotropies are imprinted during
patchy reionization \citep{santos03, zahn05a, mcquinn05}, and the
Atacama Cosmology Telescope and the South Pole Telescope are starting
to observe these
anisotropies.\footnote{http://www.hep.upenn.edu/act/act.html,
http://astro.uchicago.edu/spt/} These observations have the potential
to provide information about the duration of reionization and about
the sizes of the HII regions.

In this paper, we demonstrate that Ly$\alpha$ emitter (LAE) surveys
will likely offer the first irrefutable proof as to precisely when
reionization occurred, at least if the Universe is significantly
neutral for $z \lesssim 7$ as the high-redshift quasars suggest.  We
show that the $z = 6.6$ LAE sample from Subaru can already put limits
on $\bar{x}_i$ with just $58$ photometrically confirmed LAEs.  Surveys
on existing telescopes will also put constraints on the neutral
fraction at higher redshifts, and the James Webb Space Telescope
(JWST) -- with the appropriate observing strategy -- will effectively
be able to image HII regions during reionization.

There are several surveys that have targeted or are targeting LAEs at
epochs when the Universe may have been significantly neutral (e.g.,
\citealt{kodaira03, rhoads04, kashikawa06, cuby06, willis05,
stark07a}), and other programs will begin taking data soon (e.g.,
\citealt{casali06, mcpherson06, horton04}). These surveys take
advantage of the fact that a galaxy can produce a sizable fraction of
its flux in the Ly$\alpha$ line \citep{partridge67}. The
transmission of the Ly$\alpha$ line from a galaxy is decreased if the
neighborhood of an emitter is largely neutral, and this modulation can
be used to probe the epoch of reionization \citep{miralda98, madau00,
haiman02, santos04, furlanetto04c, furl-galaxies05, malhotra06}.

Most previous theoretical studies of high-redshift LAEs have focused
on using the evolution in the LAE luminosity function or in the
average Ly$\alpha$ line profile to constrain reionization.  The first
predictions for the impact of reionization on the LAE luminosity
function and line profile were based on the assumption that each
emitter sits in its own HII region until the final bubble percolation
stage, at which time all of the HII regions quickly merge and
reionization is completed \citep{santos04, haiman04}. In reality, the
HII regions can be much larger throughout reionization than this
simple model predicts owing to the clustering of high-redshift sources
(e.g., \citealt{sokasian03, furlanetto04a, furlanetto04b}).  When the
Universe has $\bar{x}_i = 0.5$, typically a LAE will be in an HII
region created by thousands of galaxies. Because of this clustering,
both the evolution of the LAE luminosity function and the change in
the shape of the average line profile of an emitter are a weaker
function of the neutral fraction than predicted by models based on a
single source per HII region \citep{furlanetto04c, furl-galaxies05,
mcquinn06b}.

The measured $z = 6.5$ LAE luminosity function has already been used
to constrain $\bar{x}_i$.  \citet{malhotra04} argued that the data is
consistent with no evolution in the luminosity function between $z =
5.7$ and $z = 6.5$, and this lack of evolution allowed
\citet{malhotra05} to derive the limit $\bar{x}_i \gtrsim 0.2$ at $z =
6.5$.  However, in the newest and largest LAE survey at $z > 6$,
\citet{kashikawa06} finds significant evolution between the $z = 5.7$
and the $z =6.6$ luminosity functions for luminosities $> 2\times
10^{42}$ erg s$^{-1}$, and they suggest that this evolution could be
evidence for reionization.  Furthermore, \citet{iye06} finds that the
number density of LAEs with luminosities above $1\times 10^{43}$ erg
s$^{-1}$ at $z = 7$ is $20\%-40\%$ smaller than this number density at
$z = 6.6$.

In addition to the luminosity function and the Ly$\alpha$ line
profile, the HII bubbles will influence the {\it measured} clustering
of LAEs \citep{furl-galaxies05, mcquinn06b}.  Detecting reionization
through its impact on observed clustering will be the most fool-proof
method to identify pockets of neutral hydrogen in the IGM with LAEs,
since astrophysical uncertainties and observational systematics cannot
induce large-scale correlations similar to $\sim 10 \; \Mpc$ HII
regions during reionization. We show that a LAE survey using
clustering measurements can confirm whether reionization is occurring
at $z = 6.6$ in the immediate future and that upcoming surveys will
place constraints on reionization at higher redshifts.

This paper is organized as follows.  In Section \ref{calculations}, we
describe the simulations of reionization that we employ as well as the
post-processing calculations used to generate mock Ly$\alpha$ surveys
from the simulation outputs.  These mock LAE surveys are featured in
Section \ref{LAE}.  The effect of reionization on the luminosity
function is discussed in Section \ref{lumfunc} (and on the line
profile in Appendix \ref{lineprofile}). However, the primary focus of
this work is to investigate the impact that reionization has on the
measured clustering properties of LAEs and to quantify the
detectability of reionization-induced clustering in present surveys
(Section \ref{clustering} and, for void statistics, Appendix
\ref{voids}).  We conclude in Section \ref{surveys} with a discussion
of the capabilities of upcoming LAE surveys to constrain reionization.

\section{Calculations}
\label{calculations}
\subsection{Simulations} \label{simulations}  We use two $1024^3$
N-body simulations generated with the TreePM code L-Gadget-2
\citep{springel05} to model the density field, one in a box of size
$94 \, \Mpc$ with outputs every $50$ Myr and the other in a box of
size $186\, \Mpc$ with outputs every $20$ Myr.  A Friends-of-Friends
algorithm with a linking length of $0.2$ times the mean inter-particle
spacing is used to identify halos.  We employ the higher-resolution
$94 \, \Mpc$ box to study line properties. This box is run using a
$\Lambda$CDM cosmology with $n_s = 1$, $\sigma_8 = 0.9$, $\Omega_m =
0.3$, $\Omega_{\Lambda} = 0.7$, $\Omega_b = 0.04$, and $h = 0.7$,
consistent with the first year WMAP results \citep{spergel03}.  We use
the $186 \, \Mpc$ box to study the luminosity function and clustering
properties of emitters.  This volume provides a larger sample of the
structures that are present during reionization than does the smaller
box. The $186 \, \Mpc$ box is run with a $\Lambda$CDM cosmology
updated to be more consistent with the latest WMAP results
\citep{spergel06} with $n_s = 1$, $\sigma_8 = 0.8$, $\Omega_m = 0.27$,
$\Omega_{\Lambda} = 0.73$, $\Omega_b = 0.046$, and $h = 0.7$.
The assumptions about the cosmology do not significantly affect the
morphology of reionization if we compare at fixed $\bar{x}_i$
\citep{mcquinn06b}.

The halo mass function measured from the N-body simulations matches
the \citet{sheth02} mass function for groups with 32 or more
particles.  Thirty-two particle groups correspond to a halo mass of $m
= 1\times10^9 \, \Msun$ in the $94 \, \Mpc$ box and $m = 8 \times 10^9
\, M_{\odot}$ in the $186\, \Mpc$ box. We would like to resolve halos
down to the atomic hydrogen cooling mass $m_{\rm cool}$ -- halos with
virial temperature of $10^4$ K -- which corresponds to the minimum
mass galaxy that can form stars ($m_{\rm cool} \approx 10^8 M_{\odot}$
at $z = 8$). We add unresolved halos (halos that would be comprised of
fewer than 32 particles) into the simulation using a Press-Schechter
merger tree.  This algorithm is similar to the PThalo method for
generating mock catalogs of galaxies \citep{sheth99}, and accounts
for both density and Poisson fluctuations in the abundance of halos.
We demonstrate in \citet{mcquinn06b} that this method reproduces well
the power spectrum and mass function of dark matter halos seen in
simulations.

We use an improved version of the \citet{sokasian01, sokasian03,
sokasian04} and \citet{mcquinn06b} radiative transfer code.  We have
refined the algorithm to group sources more efficiently for our work, and
this improvement is discussed in Appendix \ref{code}.  Our approach is
optimized to simulate reionization, making several justified
simplifications to speed up the computation. The code
takes the gridded density field (generated from the N-body simulation
assuming that the baryons trace the dark matter) and a list of the
ionizing sources as input, and it casts rays from each source to
compute the ionization field. We assume that the sources have a soft
UV spectrum that scales as $\nu^{-4}$ (consistent with a POPII IMF),
which is used to calculate the photo-ionization state of the gas.  The
radiative transfer code assumes perfectly sharp HII fronts, tracking
the front position at subgrid scales.\footnote{This is not true for
self-shielded regions, which can remain neutral behind the front.}
Sharp fronts are an excellent approximation for sources with a soft UV
spectrum, where the mean free path for ionizing photons is
kiloparsecs, substantially smaller than the cell size in the radiative
transfer calculations.

The radiative transfer calculations presented in this paper were
performed on a $256^3$ grid for the $94~\Mpc$ box and on a $512^3$
grid for the $186~\Mpc$ box.  The radiative transfer calculations took
between three days and two weeks on a $2.2$ GHz AMD Opteron processor.

\subsection{Ly$\alpha$ Line Calculation} \label{rt}
The shape and energetics of a Ly$\alpha$ line from a galaxy are
determined by complex processes, many of which cannot be modeled
accurately.  The Ly$\alpha$ emission depends on the amount of dust in
the galaxy, the velocity profile of the gas in the halo, and the
fraction of ionizing photons that escape from the galaxy and influence
the photo-ionization state of the nearby IGM \citep{hansen06,
dijkstra06b, santos03, tapken07, tasitsiomi06}.  Statistics that can
be used to isolate the effect that reionization has on the emitters
from all of the complicated, uncertain astrophysics are essential to
probe reionization with LAEs. In order to construct mock surveys from
which we can measure various statistics, we must make simplifying
assumptions about the Ly$\alpha$ emission that escapes from the
vicinity of a galaxy.

To calculate the emitter line profile and its transmission we employ
two schemes, a detailed, expensive method and a fast, relatively
inexpensive technique.

{\it Method 1}: We assume a Gaussian intrinsic line profile -- the
line profile that escapes from the vicinity of the galaxy -- with its
width set by the circular velocity at the virial radius of the emitter
host halo ($\sigma_{\nu} = \nu_{\alpha} \; v_{\rm vir}/c$). This
intrinsic line profile is what is commonly assumed in the literature
(e.g., \citealt{santos04} and \citealt{dijkstra06b}), but the outgoing
line can be significantly more complicated \citep{hansen06,
dijkstra06a, tapken07, tasitsiomi06}.  See the discussion in Appendix
\ref{lineprofile} for the physical motivation for the assumed line
profile. In the absence of a galactic wind, we assume the intrinsic
line is centered at Ly$\alpha$. Given the intrinsic line from an
emitter, we compute the absorption along a skewer through the
simulation box, using the density, velocity, and photo-ionization
state of the gas. The radiative transfer simulations do not compute
the temperature of the gas self-consistently.  For simplicity, we take
the gas in all regions to be at $10^4 K$.  The value of the gas
temperature and the amplitude of the sub-grid-scale density fluctuations
do not influence the amount of damping wing absorption -- the effect
that is most relevant to this study.

{\it Method 2}: First, we set the residual neutral fraction within
each HII region to zero. The residual neutral gas primarily influences
frequencies blueward of Ly$\alpha$, which this method assumes are
fully absorbed. Next, we calculate the damping wing optical depth to
absorption $\tau_{\alpha}$ along a ray through the IGM, ignoring the
effect of peculiar velocities.  We do this for only a single frequency
that starts off at $\nu' = \nu_0\, (1 -v_{\rm w}/c)$ for each source,
where $\nu_0$ is the line center and $v_{\rm w}$ is the wind velocity
(which is typically taken to be zero).\footnote{Note that the actual
wind velocity is probably closer to $0.5 \;v_{\rm w}$ such that
$v_{\rm w}$ characterizes the shift of the Ly$\alpha$ line rather than the wind
velocity.}  We then set $L_{\rm obs, \alpha} = \mathcal{T}_{\alpha} \,
L_{\alpha} \, \exp[-\tau_{\alpha}(\nu')]$, where
$\mathcal{T}_{\alpha}$ accounts for resonant absorption and is
typically between $0.1$ and $0.5$ in the absence of outflows
(\citealt{santos03} and Appendix \ref{lineprofile}). The exact value
of $\mathcal{T}_{\alpha}$ or its scaling with halo mass is not
important for our present study since $\mathcal{T}_{\alpha}$ is
degenerate with the luminosity of the emitter prior to absorption in
the IGM, $L_{\alpha}$.  This method is in contrast to Method 1 in
which for each source we calculate $\tau_{\alpha}$ at many different
frequencies.

It turns out that both methods for computing the line profile result
in very similar results.  (See Figure \ref{fig:trans} for a comparison
of the two methods.)  The agreement between the two methods occurs
because for observed emitters the virial velocity of the halo, which
sets the intrinsic width of the line in our model, is typically much
smaller than the relative velocity between the LAE and the HII front.
The agreement between the two methods justifies the use of the cheaper
method, Method 2, for computations in which the shape of the line
profile is not required.

For reference, the radius of a
bubble that produces a damping wing absorption cross section of unity
at source-frame frequency $\nu \sim \nu_0$ is \citep{loeb99}
\begin{equation}
R_b \approx 1.1 \, \bar{x}_H \,  \left(\frac{\Omega_b}{0.046}\,\frac{0.27}{\Omega_m}\right) ~{\rm pMpc} + \frac{\nu - \nu_0}{\nu_0} \, \frac{c}{H(z_g)},
\label{eqn:rbub}
\end{equation}
where pMpc denotes proper Mpc, $\bar{x}_H$ is the neutral fraction
outside of the bubble, and $z_g$ is the redshift of a LAE in the
center of the bubble.  Equation (\ref{eqn:rbub}) is correct in detail
only for a homogeneous $x_H$.  In practice, the IGM is far from
uniform and the effective $\bar{x}_H$ along each line of sight is not
equal to the globally averaged $\bar{x}_H$.  Also note that
$\tau_{\alpha}(\nu) \approx 900 \;{\rm km \, s^{-1}} \, \bar{x}_H \,
[(1+z_g)/8]^{3/2}\,[H(z_g) R_b - c \,(\nu- \nu_0)/\nu_0]^{-1}$ -- the
optical depth scales inversely with the bubble radius. (See
\citealt{miralda98} for the exact expression for $\tau_\alpha$ as a
function of the bubble size.)  If a LAE produces its own HII bubble
then the size of this bubble is
\begin{equation}
R_b = 0.15 \left(\frac{\dot{S}}{\Msun \,{\rm yr}^{-1}} \; \frac{t_{\rm age}}{10^8 \; {\rm yr}} \;\frac{f_{\rm esc}}{0.1}\right)^{\frac{1}{3}} \; \left(\frac{8}{1 +z} \right) ~ {\rm pMpc},
\label{eqn:rbauto}
\end{equation}
where $\dot{S}$ is the star formation rate, which is estimated to be
approximately $1$-$10~\Msun$ yr$^{-1}$ for observed emitters
\citep{taniguchi05}, and $t_{\rm age}$ is the lifetime of the emitter.
Equation (\ref{eqn:rbauto}) does not account for recombinations and
assumes a Salpeter IMF with $1.5 \times 10^{53} \; \dot{S}/(\Msun \,
{\rm yr}^{-1})$ ionizing photons s$^{-1}$ \citep{hui02}.  Note that
typically $\tau_{\alpha}(\nu_0) > 1$ for an emitter that sits in an
HII region created entirely by itself.  In CDM theory, high redshift
galaxies are very clustered such that the situation of a single source
creating an HII region is a rarity.

\subsection{Simple Model for Intrinsic Distribution of LAEs}
\label{simple_model} 
Given the methods discussed in Section \ref{rt} for calculating the
line profile, we also need an algorithm that tells us where the LAEs
are located to generate mock surveys.  For the analysis in this paper,
we adopt a simple model to populate the halos in the simulation with
LAEs.  We describe a mock survey of LAEs with three
parameters: 
\begin{itemize}
    \item $m_{\rm min}$ -- minimum mass halo that hosts an observed LAE
    \item $f_{\rm E}$ -- fraction of halos that host LAEs
    \item $F_{\rm c}$ -- fraction of the LAE sample contaminated by lower-redshift interlopers
\end{itemize}   
We assume that each halo in the simulation box has at most one LAE,
which is reasonable at $z > 6$ since the gas cooling time in all halos
is shorter than a Hubble time.  We also assume that $f_{\rm E}$ is not
a function of halo mass.\footnote{Even though this assumption might
not be true in detail, the majority of observed emitters reside in
halos of roughly the same mass in our model because current
observations probe the tail of the luminosity function.}  This model
is similar to those for LAEs used in \citet{dijkstra06a} and in
\citet{stark07b}.  Both \citet{dijkstra06a} and \citet{stark07b} find
that this simple parameterization is able to provide adequate fits to
the $z = 5.7$ and $z = 6.6$ luminosity functions even when fixing
$F_c$.

In the context of this model, we can write expressions for the
intrinsic number density of observed emitter candidates $\bar{n}_{\rm E} =
f_{\rm E} \, (1 + F_{\rm c}) \, \bar{n}_{\rm h}(m_{\rm min})$ as well as the
intrinsic 3-D $\bar{n}_{E}$-weighted power spectrum of emitter candidates
$P_{\rm int, E}(\bfk) = (1 -F_c)^2 \, P_{\rm hh}(\bfk, m_{\rm min}) +
1/\bar{n}_{\rm E}$.  Here, we use ``intrinsic'' to mean the value of
these functions when the effect of damping wing absorption from the
HII regions is not included.  Note that $\bar{n}_{\rm h}(m_{\rm min})$ is the
number density of halos with $m > m_{\rm min}$, $P_{\rm hh}(\bfk,
m_{\rm min})$ is the power spectrum of halos that have $m >m_{\rm
min}$ (with the shot noise component subtracted off), and this
expression for $P_{\rm int, E}$ assumes that there is no clustering of
the foreground interlopers.\footnote{For narrow band LAE surveys,
foreground interlopers come from a limited number of redshifts because
they must have a strong emission line that falls within the narrow
frequency band.  As a consequence of this redshift selection, the
clustering of foreground interlopers is more significant than if the
interlopers came from all intervening redshifts. However, unless the
contamination fraction of a LAE survey is unusually high, the extra
clustering arising from the foreground interlopers should be much
smaller than the contribution from the LAEs, justifying the
no-clustering approximation.}

We ignore peculiar velocities in our analysis.  Peculiar velocities
most significantly affect Fourier modes oriented along the line of
sight.  In narrow-band LAE surveys, the majority of the signal is
instead contributed by transverse, long wavelength modes.  Moreover,
in linear theory the peculiar velocity terms that contribute to
$P_{\rm int, E}$ are suppressed by a factor $\sim 1/b$, where $b$ is
the $\bar{n}_{E}$-weighted linear bias of emitters.  Based on the
abundance of observed $z = 6.6$ LAEs, a reasonable estimate is $b \approx 7$ in
present surveys.

This model gives the number density and clustering of emitters with $m
> m_{\rm min}$.  However, surveys do not directly infer the mass of an
emitter but instead measure its Ly$\alpha$ luminosity.  In the absence
of dust, we can map the ionizing luminosity from a source in our
simulations to its luminosity in Ly$\alpha$ photons via
\citep{osterbrock89}
\begin{equation}
L_{\alpha} =  0.67 \, E_\alpha \, (1 - f_{\rm esc}) \, \dot{N}_{\rm
ion}, \label{lyalum}
\end{equation}
where $E_\alpha = 10.2$ eV, $f_{\rm esc}$ is the fraction of ionizing
photons that escape into the IGM, and $\dot{N}_{\rm ion}$ is the
production rate of ionizing photons.  Assuming that equation
(\ref{lyalum}) holds, that $f_{\rm esc} = 0.02$, and that $L_{\rm obs,
E} = 0.5 \, L_{\alpha}$ (where the factor $0.5$ roughly accounts for
resonant absorption in the IGM), the observed Ly$\alpha$ luminosity of
the sources in our fiducial reionization model [model (i) in the next
section] is $L_{\rm obs, E} = 5 \times 10^{42} m/(10^{11} \,
M_{\odot})$ erg s$^{-1}$.  The emitters that have been observed at $z
= 6.6$ have $L_{\rm obs, E} = (2 -20) \times 10^{42}$ erg s$^{-1}$
\citep{taniguchi05}. Lowering or raising $f_{\rm esc}$ -- which is
currently unconstrained -- can decrease or increase
$L_{\rm obs, E}$ for a halo of fixed $m$.

\subsection{Three Reionization Models} \label{models} 
\begin{figure*}
\epsfig{file=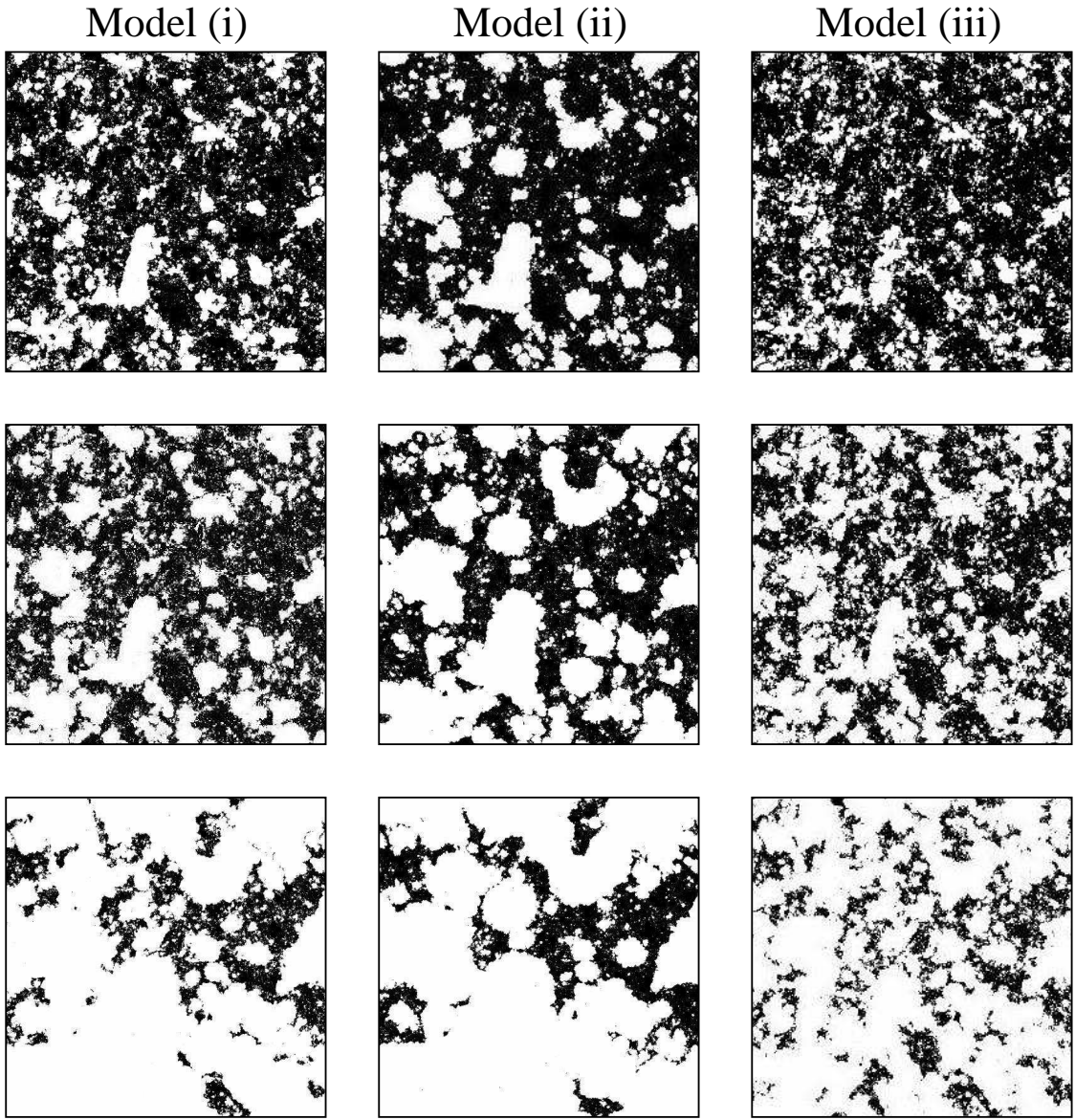, width=15cm}
\caption{Slices through the middle of the $186 \; \Mpc$ simulation box
for the three models.  The top row is at $\bar{x}_i = 0.3$, the middle
is at $\bar{x}_i = 0.5$, and the bottom is at $\bar{x}_i = 0.8$.
Model (iii), in which minihalos limit the photon mean free path, has
the smallest bubbles, whereas model (ii), in which the sources are the
most biased, has the largest. \label{fig:slice}}
\end{figure*}

We consider three models for the reionization history.  These models
bracket the possible range of morphologies for
reionization by POPII-like stars (as shown in \citealt{mcquinn06b}):

\begin{enumerate}
\item All halos above $m_{\rm cool}$ contribute ionizing photons at a
rate that is proportional to their mass $m$.  The ionizing luminosity
of source halos is $\dot{N}_{\rm ion}(m) = 3\times 10^{51} \,
[m/(10^{10} \, M_{\odot})]$ ionizing photons s$^{-1}$.  The scaling
$\dot{N}_{\rm ion} \sim m$ assumes that the star formation rate is
proportional to the amount of gas within a galaxy.  The normalization of
$\dot{N}_{\rm ion}$ is chosen such that reionization ends at $z = 7$.
Given the uncertainty in $f_{\rm esc}$ and the star formation rate in
high-redshift galaxies, the range of possible
normalizations is vast.  Fortunately, the morphology of
reionization depends very weakly on the normalization of $\dot{N}_{\rm
ion}$, as shown in \citet{mcquinn06b}.
\item Halos more massive than $m_{\rm cool}$ contribute to the
production of ionizing photons, with the ionizing luminosity of the
sources scaling as halo mass to the $5/3$ power (i.e., more massive
halos dominate the production of ionizing photons than in model (i)).
This scaling is chosen to match the relationship between star
formation efficiency and galaxy mass that is observed in low-redshift
dwarf galaxies \citep{kauffmann03} as well as the star formation rate
found in theoretical studies that include supernova feedback
\citep{dekel03, springel03}.  The total budget of ionizing photons
released is calibrated such that reionization ends at $z \approx 7$ as
in model (i).
\item Absorption by minihalos shapes the morphology of
reionization. We use the same source parametrization as in model (i)
except that the sources are twice as luminous so that reionization
ends at $z \approx 7$.  While minihalos do not contribute ionizing
photons in this model, they do act as photon sinks.  All minihalos
with $m > 10^5~M_{\odot}$ absorb incident ionizing photons out to
their virial radius until they are photo-evaporated.  This absorption
cross section is larger than the cross section found in
radiative-hydrodynamic simulations of minihalo evaporation
\citep{iliev-mh}, suggesting that this model overestimates the impact
of minihalo absorption.  The photo-evaporation timescale is roughly
the sound-crossing timescale of a halo.\footnote{We use the fitting
formula in \citet{iliev-mh} for the evaporation timescale, which
parameterizes the evaporation timescale as a function of redshift,
halo mass, and incident ionizing flux.}
\end{enumerate}

Figure \ref{fig:slice} displays slices through simulations using
models (i), (ii), and (iii).  The white regions are ionized and the
black are neutral.  Model (ii) results in the largest HII regions
because it has the most biased sources, whereas model (iii) produces
the smallest bubbles, with the maximum bubble radius restricted to be
roughly the mean free path for ionizing photons to intersect a
minihalo.

One piece of physics that is missing from these three reionization
models and that has has not been quantified in simulations of
reionization is the effect of a duty cycle for the ionizing sources on
the morphology.  It is probable that the galaxies form massive stars
and contribute ionizing photons only during certain periods.  In
Appendix \ref{duty}, we demonstrate that the duty cycle of the sources
does not affect the morphology of the ionized regions for most realistic
reionization models.

\section{LAE Maps}
\label{LAE}
\begin{figure*}
\epsfig{file=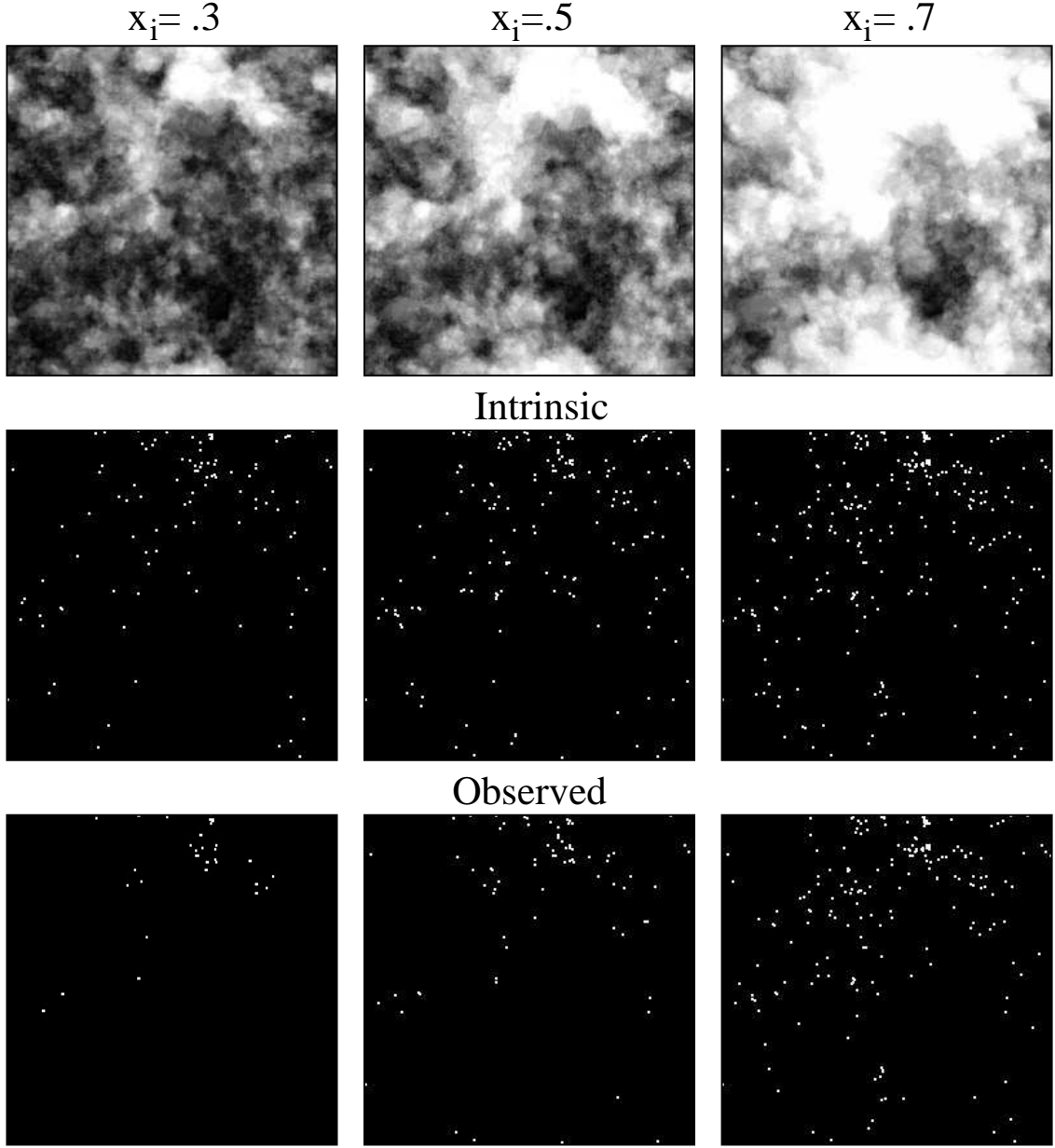, width=15cm} \caption{Top panels show the
projection of $\bar{x}_i$ in the survey volume.  In the white regions
the projection is fully ionized and in black it is neutral.  The left,
middle, and right panels are for $z=8.2$ ($\bar{x}_i = 0.3$), $z =7.7$
($\bar{x}_i= 0.5$), and $z = 7.3$ ($\bar{x}_i= 0.7$). The middle and
bottom rows are the intrinsic and observed LAE maps,
respectively, for $f_{\rm E} = 0.25$ and assuming that we can observe
unobscured emitters with $m \, \exp(-\tau_\alpha(\nu_0)) > 7\times 10^{10}
M_{\odot}$.  (Note that $L_{\rm int, E} \propto m$.)  The observed
distribution of emitters is modulated by the location of the HII
regions (compare bottom panels with corresponding top panels).  Each
panel is $94 \, \Mpc$ across (or $0.6$ degrees on the sky), roughly
the area of the current Subaru Deep Field (SDF) at $z=6.6$ \citep{kashikawa06}.
The depth of each panel is $\Delta \lambda = 130~ \AA$, which matches
the FWHM of the Subaru $9210 \, \AA$ narrow band filter. The number
densities of LAEs for the panels in the middle row are
few times larger than the number density in the SDF photometric sample
of $z=6.6$ LAEs.
\label{fig:maps}}
\end{figure*}

Figure \ref{fig:maps} shows mock LAE surveys created using snapshots
from the simulation of model (i) in the $94 ~\Mpc$ box and with a
depth of $130~ \AA$ or approximately $35 ~\Mpc$.  The dimensions of
these mock surveys are roughly the same as the $z = 6.6$ Subaru Deep
Field \citep{taniguchi05, kashikawa06}.  Each panel would subtend
$0.6$ degrees or roughly the solid angle of the moon.  These LAE maps
are generated using Method 2 in Section \ref{rt} and assuming that
$25\%$ of halos host emitters\footnote{Twenty-five percent is chosen
to match the fraction of Lyman-break galaxies at $z=3$ that meet the
selection criteria to be detected in a narrow band survey as a LAE
\citep{shapley03}, although it is probable that Lyman break galaxies
have some duty cycle as well and that the fraction of galaxies that
emit in Ly$\alpha$ evolves with redshift.} (i.e., $f_{\rm E} = 0.25$)
and that only those Ly$\alpha$ emitting halos can be observed that
have $m \, \exp(-\tau_{\alpha}(\nu_0)) > 7 \times 10^{10} ~
M_{\odot}$.  This model corresponds to setting $m_{\rm min} = 7 \times
10^{10}~\Msun$. Note that $L_{\rm int, E} \propto m$ in model (i).

The top panels in Figure \ref{fig:maps}
are the projected ionization maps for $\bar{x}_i = 0.3$ ({\it left
panel}), $0.5$ ({\it middle panel}), and $0.7$ ({\it right panel}) at
redshifts $z=8.2, ~7.7$, and $7.3$, respectively.  
The projected map is completely ionized 
in white areas and neutral in black ones.
The middle row of panels shows the intrinsic distribution of
emitters (or what would be observed if $\bar{x}_i \approx 1$), and
the bottom row of panels shows the observed distribution. (Compare the
bottom panels with their corresponding top panels to see the
reionization-induced modulation.)  The ``Intrinsic'' $\bar{n}_{\rm
E}$ in Figure \ref{fig:maps} is a few times higher then the
$\bar{n}_{\rm E}$ of the $z=6.6$ Subaru Deep Field (SDF) photometric sample
\citep{taniguchi05}.  The ``Observed'' $\bar{n}_{\rm E}$ when
$\bar{x}_i = 0.5$ in Figure \ref{fig:maps} is comparable to the
$\bar{n}_{\rm E}$ of the $z=6.6$ SDF photometric sample.  Note the
higher degree of clustering in the ``Observed'' panels
relative to the ``Intrinsic'' panels.  We
investigate the detectability of this clustering
in Section \ref{clustering}.

\begin{figure*}
 \epsfig{file=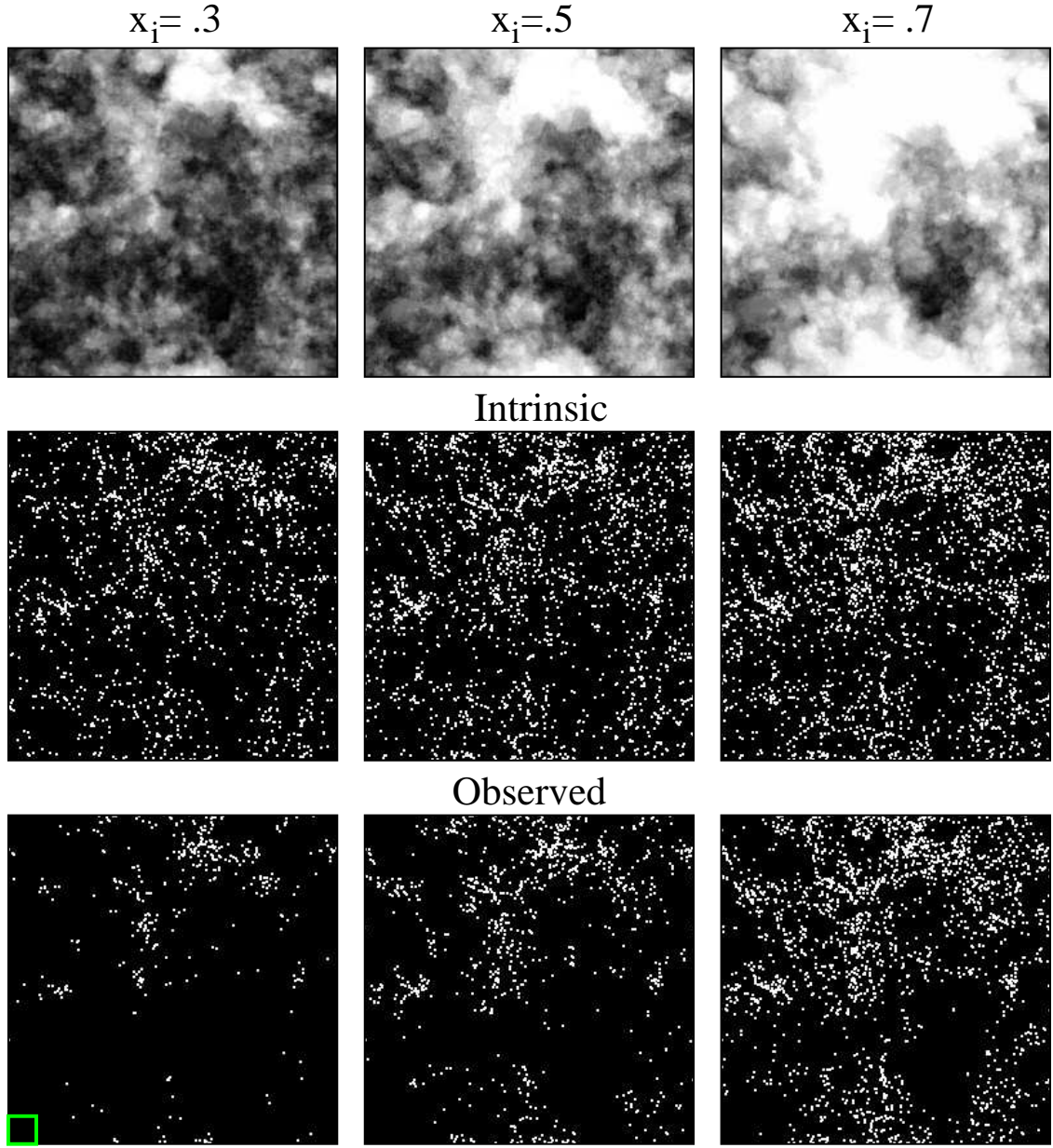, width=15cm}
\caption{Same as Figure \ref{fig:maps}, but for a futuristic LAE
survey that can detect halos down to $m \, \exp(-\tau_{\alpha}(\nu_0))
> 1\times 10^{10} M_{\odot}$ (note $L_{\rm int, E} \propto m$) and
assuming $f_{\rm E} = 0.25$.  The large-scale modulation of LAE by the
HII bubbles is clearly apparent in this survey.  The square
in the lower left-hand panel represents the $3'\times 3'$ FOV of
JWST drawn to scale.\label{fig:maps2}}
\end{figure*}

Figure \ref{fig:maps2} is the same as Figure \ref{fig:maps}, but for a
futuristic survey that is sensitive to Ly$\alpha$ emitting halos with
$m \, \exp(-\tau_\alpha(\nu_0)) > 1 \times 10^{10} ~ M_{\odot}$.
There are approximately $2000$ sources in the ``Intrinsic'' panels,
but there can be significantly fewer in the ``Observed'' panels.  The
large-scale modulation of emitters by the bubbles in Figure
\ref{fig:maps2} allows observations to image the neutral holes in the
map with the LAEs.  If the Ly$\alpha$ luminosity scales roughly with
$m$, as it does for the sources in these calculations, the mock survey
depicted in Figure \ref{fig:maps2} would entail a $\approx 50$ times
longer observation than that of Figure \ref{fig:maps} (assuming that the
observation is photon-limited). A space mission with a wide field of
view (FoV) camera could provide such a deep survey in a much shorter observation.  The James Webb
Space Telescope (JWST), set to be launched in 2013, will be able to
easily provide such a deep image.  However, the current high-redshift
program for JWST specifies a deep survey that will cover less than
one-tenth of the solid angle subtended by a single panel in Figure
\ref{fig:maps2} \citep{gardner06}.  With such a small field, it will
be difficult for JWST to probe the HII regions during reionization.
The small square in the lower left-hand panel represents the FoV of
JWST.

\section{The Luminosity Function}
\label{lumfunc} 
The luminosity function of LAEs depends sensitively on the morphology
of HII regions during reionization. If all the HII regions are smaller
than $1$ pMpc (such that $\tau_\alpha(\nu_0) > 1)$), only a small
fraction of LAEs would be observed compared to the number that would
be observed if the Universe were fully ionized.  In fact, even if the
bubbles are a few times larger than $1$ pMpc, many emitters will be
obscured.  This is because the steep, decreasing nature of the
luminosity function means that the majority of emitters have
luminosities that are within a factor $2$ of the detection threshold,
requiring for many LAEs that $\tau_{\alpha}(\nu_0)$ be significantly
less than unity to be observed.

As reionization proceeds, larger HII regions will form, allowing more
LAEs to appear out of the dark.  An extremely rapid decrease in
$\bar{n}_{\rm E}$ would be difficult to attribute to evolution in the
intrinsic properties of the LAEs rather than to reionization.  Since
the LAE luminosity function has been measured at $z =6.6$ (and
tentatively at $z=7$) and will be constrained at even higher redshifts
in the coming years, it is important to understand the signature of
reionization in the luminosity function. 

\begin{figure}
 \rotatebox{-90}{\epsfig{file=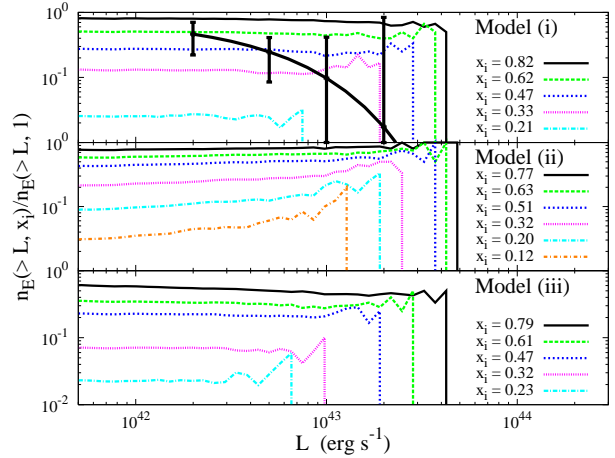, height=8.7cm}}
\caption{Ratio of the number density of LAEs with $L_{\rm obs, E}
  > L$ for different $\bar{x}_i$, $n_{\rm E}(> L, \bar{x}_i)$, to the
  number density for $\bar{x}_i \approx 1$, $n_{\rm E}(> L, 1)$.  The
  thick solid curve with the $1$-$\sigma$ errors in the top panel is
  the ratio of the $z = 6.6$ and $z = 5.7$ SDF luminosity
  functions. The effect of reionization in all models is approximately
  a uniform suppression of the luminosity function that is independent
  of luminosity.  Model (ii) has the largest bubbles such that the
  luminosity function is the least suppressed at fixed $\bar{x}_i$,
  whereas model (iii) has the smallest bubbles such that the
  luminosity function is the most suppressed. \label{fig:lumratio}}
\end{figure}

The current data on LAE luminosity functions may indicate that
reionization is happening at $z = 6.6$.  \citet{kashikawa06} finds
that there is a suppression in the bright end of their measured
luminosity function at $z=6.6$ relative to that at $z = 5.7$ at
$2$-$\sigma$ significance.  See the thick solid curve in Figure
\ref{fig:lumratio} for the ratio of the $z = 5.7$ and $z = 6.6$
best-fit LAE luminosity functions along with an estimate for the
$1$-$\sigma$ shot noise errors on this ratio \citep{shimasaku06,
kashikawa06}.\footnote{The cosmic variance errors are highly
correlated between different luminosity bins, and a $1$-$\sigma$
cosmic fluctuation will raise or suppress this ratio by $\approx 50\%$
for an ionized universe (see Section \ref{surveys}).}
\citet{kashikawa06} propose that the suppression of the high
luminosity end may imply a change in the ionization state of the IGM.
In addition, \citet{iye06} finds an additional factor of few
suppression in the luminosity function at $z = 7$ for emiters with
 $L_{\rm obs, E} > 1\times 10^{43}$ erg s$^{-1}$.

\citet{dijkstra06a} suggest that there is a more mundane explanation
for this suppression -- the evolution of the halo mass function.
Employing a similar model for LAEs to ours, they argue that the
cosmological evolution of the halo mass function between $z = 5.7$ and
$z = 6.6$ can account for the observed suppression and that it is
unnecessary to invoke reionization.  However, if \citet{dijkstra06a}
were to include dispersion in the luminosity for a given halo mass,
the high mass end of the luminosity function would be less dependent
on the evolution of the halo mass function, possibly altering this
conclusion.  Also, the model in \citet{dijkstra06a} favors $f_{\rm E}
\sim 1$ in order to fit the data at $z = 6.6$ whereas they find that
$f_{\rm E} \ll 1$ provides a better fit at $z = 5.7$.  A similar trend
for the evolution of $f_{\rm E}$ is also found in \citet{stark07b},
even though such dramatic evolution in the intrinsic properties of emitters is
not expected.

\begin{figure}
 \rotatebox{-90}{\epsfig{file=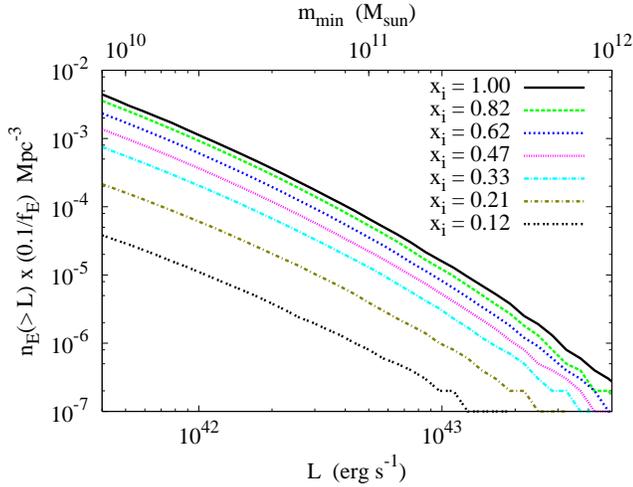, height=8.7cm}}
\caption{Number density of LAEs with $L_{\rm obs, E} > L$ at $z = 6.6$
  and for several different volume-averaged ionization fractions.
  These curves are calculated from a simulation of model (i) in the
  $186~\Mpc$ box.  The top axis shows the number density with $m \;
  \exp(-\tau_{\alpha}(\nu_0)) > m_{\rm min}$.  The mapping between $m$
  and $L_{\rm obs, E}$ is discussed in the text. \label{fig:lumfuncs}}
\end{figure}

\begin{figure}
 \rotatebox{-90}{\epsfig{file=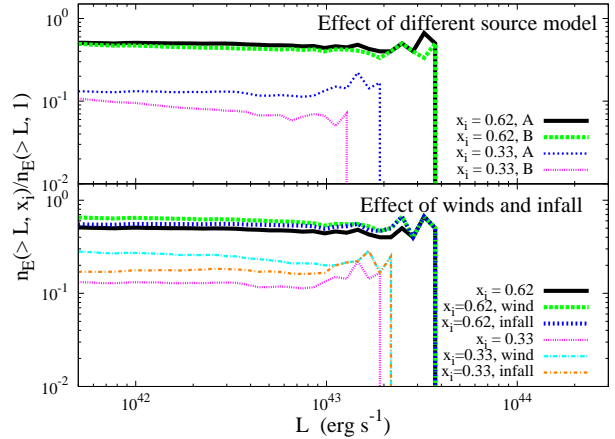, height=8.7cm}}
\caption{The same as Figure \ref{fig:lumratio}, but the curves are
computed assuming model (i).  {\it Top Panel:} Curves compare the
suppression in the fiducial model in which $L \propto m$ (source model
A) to a model in which $L$ is independent of $m$ (source model B).
{\it Bottom Panel:}  Curves labelled ``wind'' assume $v_{\rm w} =
400$ km s$^{-1}$ and labeled ``infall'' assume $v_{\rm w} =
v_{\rm vir}$. \label{fig:lumtests}}
\end{figure}

For most of the calculations in this section, we assume the simple model for
LAEs discussed in Section \ref{simple_model}.  Halos that are active
LAEs have $L_{\rm int, E}(m) = 5 \times 10^{42} \, (f_{\rm
esc}/0.02)^{-2} \, [m/ (10^{11} \,\Msun)]$ erg s$^{-1}$ in model (i).
We set $f_{\rm esc} = 0.02$ to roughly match the observed abundance of
LAEs for $f_{\rm E} = 0.1$ and $m_{\rm min} = 5 \times 10^{10}~
\Msun$.  For the other two models, we take $L_{\rm int, E}(m)$ to be
the same function as in model (i).  To be consistent with equation
(\ref{lyalum}) and the $\dot{N}_{\rm ion}$ used in models (ii) and
(iii), fixing $L_{\rm int, E}(m)$ requires adjusting $f_{\rm esc}$
slightly.

The curves in Figure \ref{fig:lumfuncs} represent the observed number
density of LAEs with luminosities that satisfy $L_{\rm
obs, E} > L$.  These curves are calculated from the $186 ~\Mpc$
simulation of model (i) at $z = 6.6$.\footnote{To perform this
calculation, we use the halo field at $z = 6.6$ in the $186$ Mpc box,
but we use the ionization field from the simulations of model (i),
which are generally at slightly higher redshifts.  This is justified
because the ionization field, when comparing at fixed $\bar{x}_i$, is
essentially independent of the redshift where this $\bar{x}_i$ is
reached \citep{mcquinn06b}.}  During reionization, the luminosity
function is suppressed by a factor that does not depend strongly on
$L$.  Figure \ref{fig:lumratio} plots the ratio of the luminosity
function at various $\bar{x}_i$ for the three models.  As with model
(i), the luminosity function in the other models is suppressed by a
largely luminosity-independent factor at each $\bar{x}_i$.  The sharp
cutoff at high luminosities in this ratio owes to the finite
simulation volume.  Model (ii) has the largest bubbles so that the
luminosity function is the least suppressed at fixed $\bar{x}_i$,
whereas model (iii) has the smallest bubbles such that the luminosity
function is the most suppressed.  

We also calculated the suppression of the luminosity function for a
wind model with $v_{\rm w} = 400$ km s$^{-1}$.  This value for $v_{\rm
w}$ is motived by the measured average velocity offset of the
Ly$\alpha$ line in strong emitters at $z \approx 3$ \citep{shapley03}.
This model probably yields an upper limit for the effect of winds at
$z >6$, where the galaxies are much less massive than at $z\approx 3$
and, therefore, not able to power such strong winds.  A wind causes a
redshift of the Ly$\alpha$ line because Ly$\alpha$ photons lose energy
when they scatter off the baryons in the wind.  Winds make the bubble
size needed for a fixed damping wing optical depth shrink.  See the
curves labeled ``wind'' in the bottom panel in Figure
\ref{fig:lumtests} and compare with the other curves with the same
$\bar{x}_i$.  This comparison illustrates the effect of a winds on the
luminosity function can be significant for $\bar{x}_i \lesssim 0.3$,
but tend to be unimportant when $\bar{x}_i \gtrsim 0.6$.  The effect
of winds on the luminosity function decreases with increasing
$\bar{x}_i$ because as the bubbles grow (increasing $\bar{x}_i$) the
ratio of the wind velocity to the Hubble flow velocity at the bubble
edge, $v_{\rm w}/(H(z_g)\;R_b)$, decreases.

We also consider a model in Figure \ref{fig:lumtests} where we set $v_{\rm w} = v_{\rm vir}(m)$,
where $v_{\rm vir}(m)$ is the halo circular velocity at the virial
radius.  This model is meant to emulate
analytic calculations in which resonant absorptions of infalling
material can obscure all wavelengths blueward of $\lambda(1 + v_{\rm
vir}(m)/c)$ \citep{santos04, dijkstra06b}.  As with winds, infall has a
relatively minor effect at $\bar{x}_i \gtrsim 0.6$, but can reduce the
suppression of the luminosity function for $\bar{x}_i \lesssim 0.3$
(Figure \ref{fig:lumtests}).

\begin{figure}
\rotatebox{-90}{\epsfig{file=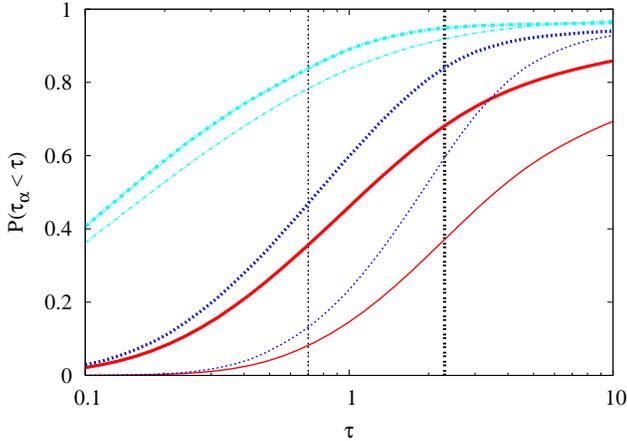, height=8.7cm}}
\caption{Probability of having $\tau_{\alpha}(\nu_0) < \tau$. The thin
curves are for LAEs with $1\times10^{10}~\Msun < m < 2 \times 10^{10}~\Msun$,
and the thick curves are for LAEs with $5\times10^{10}~\Msun < m < 1 \times
10^{11}~\Msun$.  The solid curves are for $\bar{x}_i = 0.3$, the dotted curves
are for $\bar{x}_i = 0.5$, and the dot-dashed curves are for $\bar{x}_i = 0.7$.
Assuming the observational threshold of $m > 1\times10^{10}
\;\exp(-\tau_{\alpha}(\nu_0))~\Msun$, only the emitters that
contribute to the portion of the thin (thick) curves left-ward of the
thin (thick) vertical lines are observed. \label{fig:tau_pdf}}
\end{figure}

The uniform suppression of the luminosity function was also found in
the analytic studies of \citet{furl-galaxies05}.  The explanation
provided in \citet{furl-galaxies05} for this uniform suppression is
that the most massive (most biased) sources sit in the largest
bubbles, which results in the least attenuation of their Ly$\alpha$
line.  In addition, the most massive LAEs are the most luminous (at
least in our model), requiring a larger $\tau_{\alpha}$ than an
average mass halo does to be obscured. However, at the bright end of
the luminosity function, even a slight decrease in the average
luminosity of the LAEs causes the luminosity function to change
rapidly because of its steep slope.  These effects sum to give roughly
the same suppression at the bright end as at the faint end.  In what
follows, we investigate this explanation in detail and quantify the
effect of our assumptions on our predictions for the evolution of the
luminosity function.

Figure \ref{fig:tau_pdf} illustrates the dependence of damping wing
absorption on halo mass, plotting the cumulative probability
distribution of $\tau_{\alpha}(\nu_0)$, $P(\tau_{\alpha}(\nu_0) <
\tau)$.  The thick curves are $P(\tau_{\alpha}(\nu_0) < \tau)$ for
LAEs with $5\times10^{10}~\Msun < m < 1 \times 10^{11}~\Msun$, and the
thin ones are $P(\tau_{\alpha}(\nu_0) < \tau)$ for LAEs with
$1\times10^{10}~\Msun < m < 2 \times 10^{10}~\Msun$.  This figure
demonstrates that the most massive LAEs have smaller
$\tau_{\alpha}$ because they sit in larger bubbles than less massive
LAEs.  This effect decreases with increasing $\bar{x}_i$ because the
bubbles become less biased as they grow. ({Compare the difference
between the two solid curves -- $\bar{x}_i = 0.3$ -- to the difference
between the two dot-dashed curves -- $\bar{x}_i = 0.7$.})  In
addition, the most massive LAEs are the most luminous in our model,
which requires a larger $\tau_{\alpha}$ than less massive LAEs do for
their Ly$\alpha$ luminosity to fall below the survey threshold $L_{\rm min}$.

Figure \ref{fig:tau_pdf} suggests that our predictions for the
luminosity function depend on the scaling of the LAE bias with
luminosity.  This is worrisome because this scaling is very uncertain.
How dependent are our predictions on the simple mapping we assume
between halo mass and Ly$\alpha$ luminosity?  One plausible
extreme LAE model is, rather than $L_{\rm int, E} \propto m$,
for  $L_{\rm int, E}$ to be independent of halo mass.  To achieve this,
we kept the same halo positions but randomized the luminosities of the
halos while maintaining the same luminosity function as in the fiducial Ly$\alpha$ emitter model.  We refer to
this source model as model B and the fiducial model as model A.  The
top panel in Figure \ref{fig:lumtests} compares the suppression of the
luminosity function in these two models. The suppression in the two
models is similar, particularly for the $\bar{x}_i = 0.62$ case. The
agreement between these two models suggest that our predictions are
not strongly dependent on our prescription for the luminosity
function.

When $\bar{x}_i = 0.33$, the luminosity function is more suppressed at
the high mass end for source model B than for source model A,
imparting some scale dependence in the suppression for model B.  The
scale dependence induced by model B can be thought of as the maximum
scale dependence that can be imparted by reionization since bias is
uncorrelated with luminosity in this model. Therefore, it is difficult
for reionization to be solely responsible for the scale-dependent
suppression that may have been observed in
\citet{kashikawa06}.\footnote{It is possible to induce more of a scale
dependence in the suppression by making the luminosity function steeper
than we have assumed at large luminosities.  However, we find that in
practice a steeper bright-end luminosity function does not result in
much additional suppression.  This result owes to the large dispersion in
$\tau_{\alpha}$ for halos of fixed $L_{\rm obs, E}$.}

If a LAE survey suffers from significant contamination or
incompleteness, these systematics will affect the normalization and
shape of the luminosity function, complicating any inference that the
evolution is due to neutral regions.  Also, intrinsic evolution in the
source properties with redshift could be difficult to distinguish from
reionization.  If evolution in the observed luminosity function indeed
owes to reionization, then reionization will also increase the
clustering of observed emitters, whereas systematic effects and
intrinsic evolution cannot change the clustering in the same manner.
The impact of reionization on clustering is discussed in the following
section.

\section{Clustering}
\label{clustering}

Suppose that upcoming observations confirm the substantial decrease in
$\bar{n}_{\rm E}$ between $z = 5.7$ and $z = 6.6$ found in
\citet{kashikawa06}.  This evolution could be explained in three ways (or
some combination thereof):
\begin{enumerate}
\item {\it Decreasing Duty Cycle}: The number density of halos that host
emitters is decreasing, but the observed emitters sit
in halos of the same mass at $z= 6.6$ as at $z = 5.7$. 
\item {\it Increasing Halo Mass}: The average halo mass of LAEs is
increasing between $z = 5.7$ and $z = 6.6$.
\item {\it Reionization}: Patchy reionization is still occurring at $z
 = 6.6$.  Neutral regions are obstructing the line of sight to some
 emitters, decreasing their observed abundance.
\end{enumerate}
In case (i), the clustering properties of the sources will be
essentially unchanged in the $150$ million years between $z = 5.7$ and
$z = 6.6$.  Only in cases (ii) and (iii) can the amount of clustering
increase significantly. If a high-redshift LAE survey is able to
distinguish case (ii) from case (iii), it will be capable of
determining whether reionization is happening at $z = 6.6$.

\begin{figure}
  \rotatebox{-90}{\epsfig{file=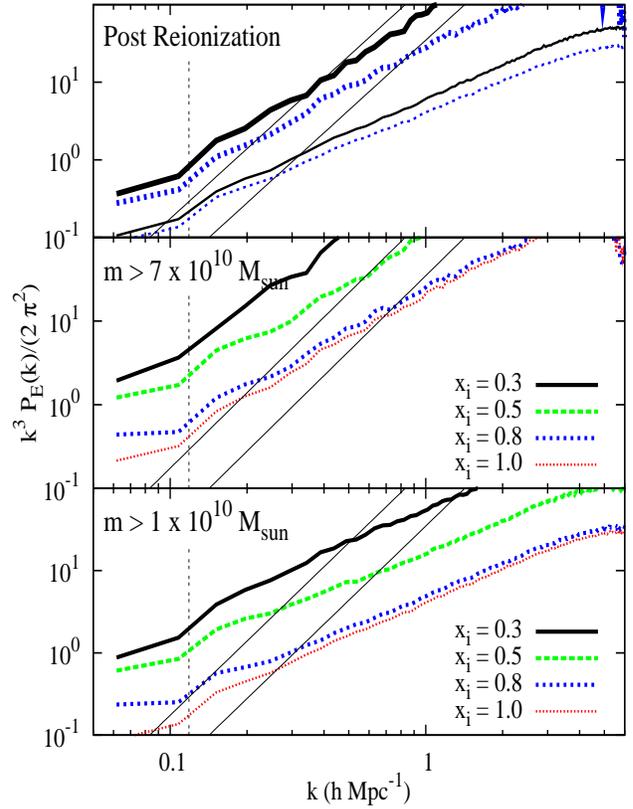, height=8.5cm,
  width=11cm}}
\caption{Dimensionless 3-D power spectrum of $\delta_{\rm E} \equiv
{n}_{\rm E}/\bar{n}_{\rm E} - 1$, calculated using the $186$ Mpc
simulation of model (i).  {\it Bottom Panel}: The power spectrum of
emitters with $m \, \exp(-\tau_{\alpha}(\nu_0)) > 1 \times 10^{10}~
M_{\odot}$.  All curves are from redshifts between $z = 6.9$ and $z =
8.3$.  {\it Middle Panel}: The power spectrum of emitters with $m \,
\exp(-\tau_{\alpha}(\nu_0)) > 7 \times 10^{10}~ M_{\odot}$, and
$\bar{n}_{\rm int, E} \approx 10^{-3} \, f_{\rm E} \; \Mpc^{-3}$.
{\it Top Panel}: The intrinsic clustering of emitters (or,
equivalently, the clustering when $\bar{x}_i \approx 1$).  The thick
curves are if emitters can be observed in halos with $m > 1\times
10^{11}~\Msun$ at $z = 6.9$ ({\it dashed curve}, $\bar{n}_{\rm E} = 1
\times 10^{-4} \, f_{\rm E} \; \Mpc^{-3}$) and at $z = 8.0$ ({\it
solid curve}, $\bar{n}_{\rm E} = 5\times 10^{-5} \, f_{\rm E} \;
\Mpc^{-3}$).  The thin solid and dashed curves are the same but in
halos with $m > 1\times 10^{10}~ \Msun$ (with $\bar{n}_{\rm E} = 0.03
\, f_{\rm E} \, \Mpc^{-3}$ at $z = 6.9$ and $\bar{n}_{\rm int, E} =
0.02 \, f_{\rm E} \; \Mpc^{-3}$ at $z = 8$, respectively).  {\it All
Panels}: The straight, vertical line corresponds to $k =
2 \pi\, h \, R^{-1}$, where $R = 80~\Mpc$ -- roughly the angular
extent of the $z = 6.6$ LAE survey in the SDF.  The diagonal solid
lines are the shot-noise power spectrum for $\bar{n}_{\rm E} = 1
\times 10^{-4} ~\Mpc^{-3}$ and $\bar{n}_{\rm E} = 5\times 10^{-4}
~\Mpc^{-3}$.  The former $\bar{n}_{\rm E}$ is approximately the value of
$\bar{n}_{\rm E}$ in the SDF photometric sample.
\label{fig:clustering}}
\end{figure}

\begin{figure}
  \rotatebox{-90}{\epsfig{file=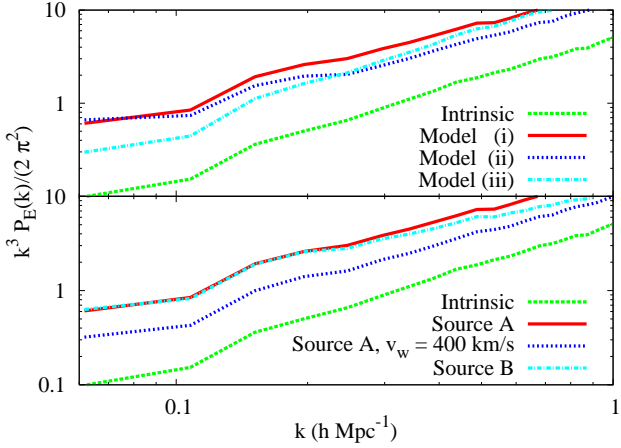, height=8.8cm}}
\caption{{\it Top Panel:} Dimensionless 3-D power spectrum of
$\delta_{\rm E}$ for three different models of reionization,
calculated from snapshots that have $\bar{x}_i = 0.5$ and assuming $m
\, \exp(-\tau_{\alpha}(\nu_0)) > 1 \times 10^{10}~ M_{\odot}$.  In
addition, a curve for the intrinsic (or $\bar{x}_i \approx 1$) power
spectrum is included for comparison. {\it Bottom Panel}: Comparison of
the $\bar{x}_i \approx 0.5$ LAE power spectrum of source model A (in
which $L_{\rm int, E} \sim m$) to both the LAE power spectrum of a
wind model that uses source model A and $v_{\rm w} = 400$ km s$^{-1}$
and to the LAE power spectrum of source model B (in which $L_{\rm int,
E}$ is independent of $m$). The curve labeled ``intrinsic'' uses source
model A, and all curves in the bottom panel use reionization model
(i).
\label{fig:clustering_models}}
\end{figure}

The curves in Figure \ref{fig:clustering} are the 3-D power spectrum
of $\delta_{\rm E} \equiv {n}_{\rm E}/\bar{n}_{\rm E} - 1$, calculated from
a simulation of model (i).  The shot noise component of this power
spectrum has been removed from these curves. Note that these curves
depend only on $m_{\rm min}$ and the ionization field; they do not
depend on $f_{\rm E}$. We plot the 3-D rather than the 2-D power
spectrum because the 3-D power spectrum makes use of all the two-point
information that is available in our simulation volume, minimizing
cosmic variance. Note that all of our conclusions about the effect of
reionization on the LAEs would be the same if we considered the 2-D
power spectrum.  The curves in the middle panel in Figure
\ref{fig:clustering} represent the case that the observed LAEs have $m
\, \exp(-\tau_{\alpha}(\nu_0)) > 7 \times 10^{10}~ M_{\odot}$. (Note
that $L_{\rm int, E} \sim m$.)  This threshold along with $f_{\rm E}
\approx 0.1$ yields the $\bar{n}_{\rm E}$ measured in the SDF at $z =
6.6$.

The curves in the bottom panel in Figure \ref{fig:clustering}
represent the case in which LAEs are observed in less massive halos
than the middle panel, halos with $m \, \exp(-\tau_{\alpha}(\nu_0)) >
1\times 10^{10}~ M_{\odot}$.  The curves correspond to the signal
for surveys that are more sensitive than Subaru such as JWST (unless
$f_{\rm E} \lesssim 0.01$).  The LAE power spectrum at fixed $\bar{x}_i$
has a slightly lower amplitude when computed from more abundant halos.

During reionization, the power spectrum of LAE fluctuations changes
rapidly.  Between $\bar{x}_i \approx 1$ (when $z = 6.7$ in the
simulation of model (i)) and $\bar{x}_i = 0.5$ (when $z = 7.5$), the
amplitude of the power spectrum increases by roughly a factor of
$\approx 3-4$ ({see middle panel in Fig. \ref{fig:clustering}}).  The
amplitude increases by another factor of $\approx 2$ by $\bar{x}_i =
0.3$ (when $z = 8$).  If LAE surveys detect a rapid increase in the
amplitude of the power spectrum with redshift, it would be difficult
to attribute this to anything other than reionization.

The straight, solid diagonal lines in Figure \ref{fig:clustering}
represent the shot-noise power spectrum for $\bar{n}_{\rm E} = 1
\times 10^{-4} ~\Mpc^{-3}$ and $\bar{n}_{\rm E} = 5\times 10^{-4}
~\Mpc^{-3}$.  The former $\bar{n}_{\rm E}$ is approximately the
$\bar{n}_{\rm E}$ measured in the SDF photometric sample
\citep{kashikawa06}.  Notice that for a survey with either value of
$\bar{n}_{\rm E}$, clustering can be detected on large scales provided
that the survey volume is large enough (i.e., the shot-noise line is
below the other curves).  The vertical dashed lines in Figure
\ref{fig:clustering} correspond to $k = 2 \pi\, h \,R^{-1}$, where
$R = 80~\Mpc$ -- roughly the angular extent of the $z = 6.6$ LAE
survey in the SDF.  These curves imply that if reionization is
happening at $z = 6.6$, fluctuations in the LAE field can be imaged
even in current programs.

The results in the bottom two panels of Figure \ref{fig:clustering}
should be contrasted with what observations would see if, instead of
the increased clustering owing to reionization, it was enhanced by
evolution in the intrinsic properties of the LAEs.  A similar
evolution in the clustering properties to the evolution caused by
reionization could conceivably be produced if the bias of the sources
were to increase with redshift -- i.e. if the LAEs sit in more massive
halos at $z = 6.6$ than at $z =5.7$.

In the top panel in Figure \ref{fig:clustering}, the thin curves have
a $10$ times smaller $m_{\rm min}$ than do the thick ones.  For fixed
$f_{\rm E}$, there are over $100$ times more emitters in the survey from
which the thin curves are calculated compared to the survey from which
the thick curves are computed.  A change in $m_{\rm min}$ that leads
to a change in $\bar{n}_{\rm E}$ by a factor of over $100$ yields a smaller
variation in $P_{\rm E}$ than the change in $P_{\rm E}$ between when
$\bar{x}_i \approx 1$ and when $\bar{x}_i = 0.5$, even though
$\bar{n}_{\rm E}$ differs by only a factor of $3$ between these curves
(fixing $f_{\rm E}$). Reionization causes the clustering
of LAEs to evolve much more quickly than is possible with intrinsic
evolution.

Let us develop a general understanding of the magnitude by which
intrinsic evolution of the LAEs can change the amount of clustering.
On large scales, $P_{\rm E} = b^2 \, P_{\rm DM}$, where $b$ is the
intrinsic large-scale bias and $P_{\rm DM}$ is the dark matter power
spectrum.  The large-scale bias calculated using Press-Schechter
theory at $z = 7$ is $b^2 = 28$ for $m_{\rm min} = 1\times
10^{10}~\Msun$, $b^2 = 58$ for $m_{\rm min} = 5\times 10^{10}~\Msun$,
$b^2 = 65$ for $m_{\rm min} = 7\times 10^{10}~\Msun$, and $b^2 = 77$
for $m_{\rm min} = 1\times 10^{11}~\Msun$.  \emph{The ratio of the $P_{\rm E}$ for $\bar{x}_i \approx
1$ and for different $m_{\rm min}$ is typically smaller than the ratio
between $P_{\rm E}$ for models in which $\bar{x}_i$ differs by $\Delta
\bar{x}_i \approx 0.3$ .}

The top panel in Figure \ref{fig:clustering_models} compares the power
spectrum of $\delta_{\rm E}$ for reionization models (i), (ii), and
(iii) and for $\bar{x}_i = 0.5$ with $m_{\rm min} =
1\times10^{10}~\Msun$.  Model (iii) has the least power on large
scales -- despite having the fewest observed emitters of the three
models -- because it has the smallest bubbles.  However, the
differences between the curves for $\bar{x}_i = 0.5$ are not as
substantial as the differences between $\bar{x}_i = 0.3$, $\bar{x}_i =
0.5$, and $\bar{x}_i \approx 1$.  Therefore, it will be easier to
constrain $\bar{x}_i$ to $\sim 30\%$ with emitter clustering than it
will be to constrain the details of the reionization process.

The bottom panel in Figure \ref{fig:clustering_models} compares the
power spectrum of $\delta_{\rm E}$ computed from a LAE field that uses
the fiducial source model (model A) and from a LAE field that uses
source model B, which was discussed in Section \ref{lumfunc}, where
luminosity is independent of halo mass for all halos above $m_{\rm
min}$.  All curves are calculated with $m_{\rm min} =
1\times10^{10}~\Msun$.  The bottom panel in Figure
\ref{fig:clustering_models} also plots the power spectrum for a LAE
model with $v_{\rm w} = 400$ km s$^{-1}$ for all emitters.  Winds suppress the amount of clustering.

If a LAE survey detects excess power on large scales, a skeptic might
contend that this is caused by a rare large-scale structure in the
survey field rather than by reionization.  Fortunately, there is a
simple test that may help to distinguish between these two hypotheses:
Survey the field using a different selection criterion, such as
H$\alpha$ emission or by the Lyman-break technique.  If the
reionization hypothesis is correct, less clustering is expected in the
second survey, and if we cross correlate with the Ly$\alpha$ survey,
again the same excess clustering will be present.\footnote{There is a
second-order effect owing to Jeans mass suppression in ionized regions
inhibiting galaxies with $m \lesssim 10^9~ M_{\odot}$ from forming.  A
high redshift galaxy would be composed of many of these smaller mass
galaxies owing to past merger events, and therefore its Ly$\alpha$
emission might be influenced by the local reionization history
\citep{babich06, pritchard07}.  This effect may result in reionization
also influencing the clustering of all galaxies, not just LAEs.
Quantifying the relevance of this effect is difficult, but it is
certainly much smaller than the modulation we consider.}

This test does not rule out the possibility that the LAEs are
intrinsically more biased than the galaxies selected with the other
selection criterion.  This possibility may be surprising because at
lower redshifts LAEs are associated with younger, less biased systems
than the Lyman break galaxies.  However, let us suppose that this
circumstance is the case. On scales where the intrinsic bias is linear
and shot noise is unimportant, the Fourier space fluctuations of the
two galaxy fields should have the same phase at each $\bfk$ if
$\bar{x}_i \approx 1$.  If this is not true, then this would be
evidence for reionization.  Of course, on large enough scales the
emitter field, $\delta_{\rm E}$, will have the same phase as the
galaxy field, $\delta_g$, even during reionization, but the presence
of HII bubbles will extend the range of scales over which the phases
do not agree.  The cross correlation coefficient $r(k) = \langle
\delta_{{\rm E}, k} \, \delta_{g, k} \rangle /(\langle |\delta_{{\rm
E}, k}|^2 \rangle \; \langle |\delta_{g, k}^2|^2 \rangle)^{0.5}$
compares the phases of two fields.  If $\bar{x}_i = 0.5$ and $m_{\rm
min} = 5\times 10^{10}~\Msun$, we find in model (i) that $r \approx
0.9$ at $k = 0.1~\Mpc^{-1}$, whereas if $\bar{x}_i \approx 1$ then $r
\approx 1$ at $k = 0.1~\Mpc^{-1}$ (even if the $m_{\rm min}$ differ by
a factor of a few between the $\delta_{\rm E}$ and $\delta_g$ fields).

Another potential test to check whether the fluctuations in the LAE
field owe in part to reionization is to note that the HII regions
break the rotational isotropy of the LAE field because Ly$\alpha$
absorption depends on the ionization state of the IGM in only the
line-of-sight direction.\footnote{Physically, a uniform distribution of LAEs
and a spherical HII region will result in a non-spherical observed
distribution of LAEs within the HII region, with the distribution of
LAEs truncated at $\sim 1$ pMpc from the edge of the HII region along
the line of sight direction.  In reality, the distribution of emitters
and the shape of HII regions is more complicated than in this
spherical example, which reduces the magnitude of this effect.}  We have
investigated the statistic $\langle |\delta_{\rm E}(\bfk \cdot
\nhat)|^2 \rangle$ for different orientations of $\nhat$ and have
found no significant signature of angular anisotropy in our simulation cube
(neglecting redshift-space distortions owing to the peculiar velocity
field).

In this section, we focused on the $\bar{n}_{\rm E}$-weighted 3-D
power spectrum.  The reader might wonder why we did not use the
luminosity-weighted power spectrum (or even some more general
weighting) instead, which can contain more information about
reionization.  We find that a luminosity-weighted power spectrum
results in a similar spectrum of fluctuations.  The reason for this is
that most emitters in a survey are near the detection threshold
because of the decreasing nature of the mass function.  It only takes
minor attenuation for most LAEs to not be observed, such that the
major source of fluctuations derives from whether an emitter is
detected or not.

\subsection{Factors that Shape the LAE Power Spectrum}
\label{analytic_model}
\begin{figure}
  \rotatebox{-90}{\epsfig{file=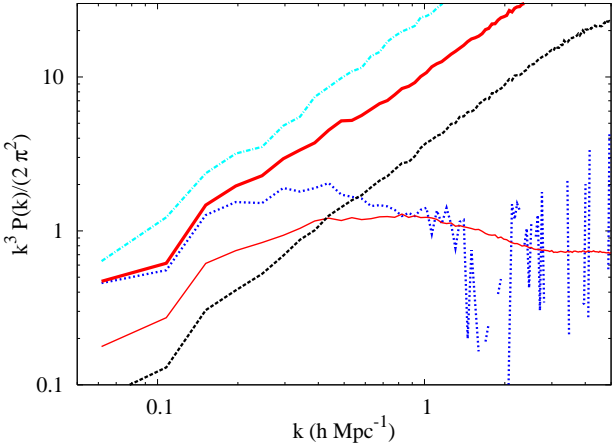, height=8.8cm}}
\caption{This plot investigates the effects that shape $k^3 P_{\rm
E}(k)/2\pi^2$.  The dashed curve is the intrinsic LAE power spectrum
for $m_{\rm min} = 1\times 10^{10}~\Msun$.  The thick solid curve is
$k^3 P_{\rm E}(k)/2\pi^2$ for $\bar{x}_i = 0.5$ and $m_{\rm min} =
1\times 10^{10}~\Msun$.  The dotted curve is the same as the thick
solid except that each of the halos has been randomly displaced in the
box prior to computing the observed LAE field.  The dot-dashed curve
is the same as the thick solid except that the halo field is uniformly
translated by some arbitrary distance relative to the ionization field
(to make $\bar{x}_i$ and $\delta_g$ uncorrelated) prior to computing
the observed LAE field.  The shot noise contribution has been
subtracted from the curves that represent $P_{\rm E}$.  The thin solid
curve is five times the power spectrum of $\delta_{x_i}$, included to
compare its spectral dependence with the other curves.
\label{fig:understand}}
\end{figure}

Let us develop a simple model for the LAEs to understand the reason
that reionization has such a large effect on $P_{\rm E}$.  We will not
use this model to predict $P_{\rm E}$, but rather to provide a
framework with which to understand the different effects that
contribute to $P_{\rm E}$ during reionization.  For this model, we
assume that only emitters in bubbles of radius $l$ are observed, where
$l \sim 1$ pMpc.  In reality, $l$ depends on the emitter luminosity
and on the minimum luminosity of the survey.  We can write the
observed number density of LAEs at position $\bfx$ as $n_{\rm E}(\bfx)
= n_g(\bfx) \, {\tilde{x}_i}^\gamma(\bfx)$, where $\tilde{x}_i$ is the
ionization field smoothed over a sphere of radius $l$, and $n_g$ is
the local number density in galaxies that emit in Ly$\alpha$.  If a
bubble is defined as a fully ionized sphere then $\gamma = \infty$.
However, in practice regions are never fully ionized, and for our
purposes it suffices to leave $\gamma$ as a free parameter.  

In this model, the correlation function of LAEs can be written as (omitting
constants)
\begin{equation}
\langle
\delta_{\rm E}\, {\delta_{\rm E}}' \rangle =  X^2\; \left(\langle {\tilde{x}_i}^\gamma \,{\tilde{x}_i}^{\gamma'}\rangle + 2\;\langle {\tilde{x}_i}^\gamma \delta_g \,{\tilde{x}_i}^{\gamma'}\rangle + \langle {\tilde{x}_i}^\gamma \delta_g \,{\tilde{x}_i}^{\gamma'} {\delta_g}' \rangle \right),
\label{eqn:corrfunc}
\end{equation}
where $\delta_{\rm E}$ and $\delta_{g}$ are the overdensities in
observed LAEs and Ly$\alpha$ emitting galaxies, respectively, and
where $X = \bar{n}_g/\bar{n}_{\rm E}$.  Even though this model is
simplistic, if we had the full model for this correlation function it
would have a similar decomposition to the decomposition seen in the
right-hand side (RHS) of equation (\ref{eqn:corrfunc}).

Which terms on the RHS of equation (\ref{eqn:corrfunc}) shape
$\delta_{\rm E}$?  The first term is generated only by the bubbles.
On scales where this term is dominant, $P_{\rm E}$ is independent of
the intrinsic clustering of the LAEs.  To investigate the importance
of this term, we randomly displaced all the halos in the simulation
box such that the intrinsic halo field is Poissonian prior to
computing the LAE field, $\delta_{\rm E}(\bfx)$.  This operation makes
the second and third terms in equation (\ref{eqn:corrfunc}) zero for
finite separations.  The power spectrum of $\delta_{\rm E}$ using the
displaced halo field for $\bar{x}_i = 0.5$ is given by the dotted
curve in Figure \ref{fig:understand} (with the shot noise contribution
subtracted off).  Compare this curve to the thick solid curve, which
represents $P_{\rm E}$ computed from the true halo field.  On large
scales, these curves agree fairly well, implying that the first term
on the RHS of equation (\ref{eqn:corrfunc}) is important at tens of
Mpc scales and greater.  This result qualitatively agrees with the
analytic study of \citet{furl-galaxies05}, which found that the
bubbles dominate the large-scale emitter clustering.

The thin solid curve in Figure \ref{fig:understand} is five times the
power spectrum of $\delta_{x_i}$ (the normalization is chosen to
facilitate a comparison of the shape of this curve with the shape of
other curves).  As expected, the spectral shape of the thin solid and
dotted curves are similar.  However, unlike the thin solid curve, the
dotted curve is consistent with zero at $k \gtrsim 2 \; h
\;\Mpc^{-1}$.  This difference is because sub-pMpc features in the
bubbles do not affect the LAE field.

We have quantified the importance of the first term, but what about the
second and third terms on the RHS of equation (\ref{eqn:corrfunc})?
These terms depend in part on the covariance between $\delta_g$ and $x_i$.
To investigate the importance this covariance, we translated the halo
field in the simulation box relative to the ionization field such that
$\delta_g$ and $x_i$ become uncorrelated. $P_{\rm E}$ increases at all
$k$ by $\approx 50 \%$ from this operation ({compare dot-dashed curve
with thick solid curve in Fig. \ref{fig:understand}}), and the shape
of $P_{\rm E}$ is maintained.

Since $P_{\rm E}$ is not drastically changed by correlations between
$\delta_g$ and $x_i$, this result motivates the assumption that
$\delta_g$ and $x_i$ are uncorrelated to understand $P_{\rm E}$.  The
second term on the RHS of equation (\ref{eqn:corrfunc}) is zero with
this assumption and the third term becomes $X^2 \, \langle
\delta_{g}\, {\delta_{g}}' \rangle \; \langle {\tilde{x}_i}^\gamma
\,{\tilde{x}_i}^{\gamma'} \rangle$.  With these simplifications, the
Fourier transform of equation (\ref{eqn:corrfunc}) becomes
\begin{eqnarray}
  P_{\rm E}(\bfk) &=&  X^2 ~\biggl[  P_{\tilde{x}^\gamma \tilde{x}^\gamma}(\bfk)\nonumber \\
  & &+  \int d^3\tilde{\bfk}' \;P_{gg}(\bfk - \bfk')\, P_{\tilde{x}^\gamma \tilde{x}^\gamma}(\bfk') \biggr],
\label{eqn:pklae}
\end{eqnarray} 
where $d^3\tilde{\bfk} = (2\pi)^{-3} \;d^3{\bfk}$.

At $k \gtrsim 1~h\, \Mpc^{-1}$, $P_{\rm E}$ is a constant times the
intrinsic power spectrum ({see the curves in
Fig. \ref{fig:clustering}}).  This limit is easy to understand in this
model.  On these scales, $P_{\tilde{x}^\gamma \, \tilde{x}^\gamma}
\approx 0$ and $P_{\rm E} \approx X^2 \int d^3\tilde{\bfk}'
P_{gg}(\bfk - \bfk')\, P_{\tilde{x}^\gamma \tilde{x}^\gamma}(\bfk')
\approx X^2 \,P_{gg}(k) \int d^3\tilde{\bfk}'\, P_{\tilde{x}^\gamma
\tilde{x}^\gamma}(\bfk')$.  The $\tilde{x}_i^{\gamma}$ field is
primarily composed of zeros and ones, such that $\int d^3
\tilde{\bfk}'\, P_{\tilde{x}^\gamma \tilde{x}^\gamma}(\bfk')
=\langle\tilde{x}^\gamma \, {\tilde{x}^{\gamma'}}\rangle(0) \approx
\langle\tilde{x}^\gamma\rangle = X^{-1}$, where
$\langle\tilde{x}^\gamma \, {\tilde{x}^{\gamma'}}\rangle(0)$ is the
correlation function evaluated at zero separation. Therefore, equation
(\ref{eqn:pklae}) becomes $P_{\rm E} \approx X\, P_{gg}$.  At
$\bar{x}_i = 0.5$, uncorrelated $\delta_g$ and $x_i$ results in $X
\approx 10$ in our calculations, and $10$ yields the small-scale
increase between $P_{\rm gg}$ and $P_{\rm E}$ in
Figure \ref{fig:understand} ({compare dashed and dot-dashed curves}).

If $P_{\rm E} \propto P_{gg}$ on small scales then the proportionality
constant must be $X$ to give the correct shot noise term.  In model
(i), when $\bar{x}_i = 0.5$ then $X \approx 3$, and when $\bar{x}_i =
0.3$ then $X \approx 8$.  Interestingly, these factors yield roughly
the small scale power increase we see in Figure \ref{fig:clustering}.
Therefore, the relation $P_{\rm E} \approx X\, P_{gg}$ at $k \gtrsim
1~\Mpc^{-1}$ seems to hold in general and provides a consistency check
for the reionization hypothesis, where $X$ can potentially be
estimated using a slightly lower redshift LAE survey to
derive $\bar{n}_g$.

\subsection{Detectability}
\label{detectability}

\begin{figure}
  \rotatebox{-90}{\epsfig{file=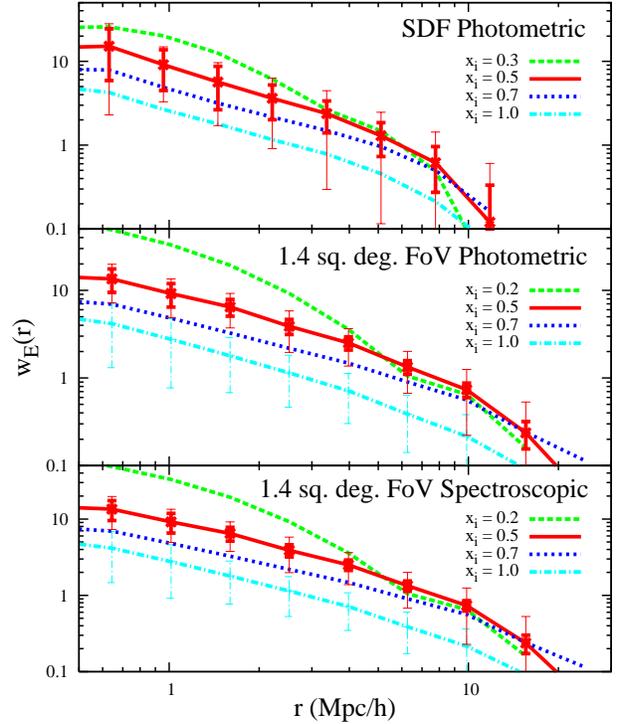, height=8.2cm}}
\caption{Angular correlation function of emitters at $z = 6.6$,
assuming that observed emitters reside in halos with $m\;
\exp(-\tau_{\alpha}(\nu_0)) > 7 \times 10^{10} ~\Msun$.  The curves in
the top panel are calculated in the same volume and with the same
number of emitters, $58$, as the SDF photometric sample.  The bottom
two panels are in a volume a slightly larger volume than the upcoming
$1$ sq. deg. Subaru/XMM-Newton Deep Survey (SXDS), with $250$ emitters
in the middle panel and with $190$ in the bottom one.
The thick error bars owe to shot noise, and the thin owe to shot noise
plus cosmic variance.  To calculate these errors, we conservatively
assume $F_c = 0.25$ in the top two panels ($F_c = 0$ in the bottom
panel).  Current surveys can potentially distinguish an ionized
universe (the curves labeled ``intrinsic'') from a universe with
$\bar{x}_i \lesssim 0.5$.
\label{fig:wcorr}}
\end{figure}

\begin{figure}
{\epsfig{file=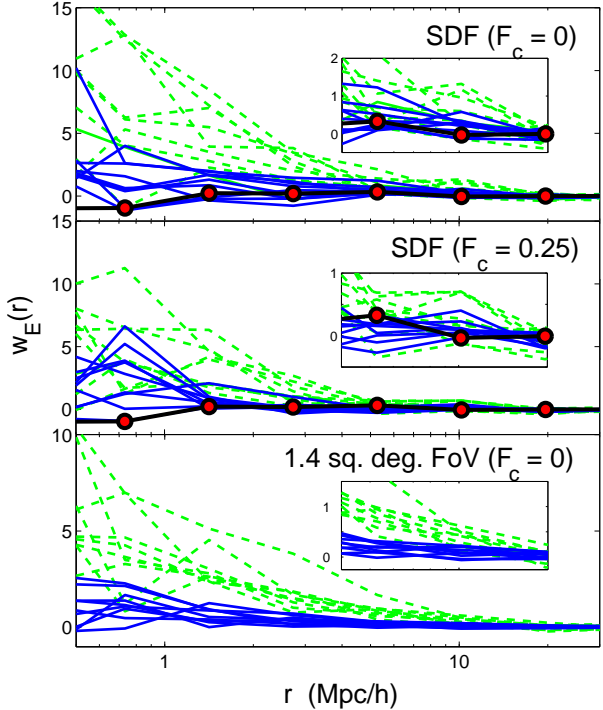, height=10cm}}
\caption{Angular correlation function of emitters at $z = 6.6$
measured from independent volumes in the $186~\Mpc$ simulation box of
model (i) for $\bar{x}_i \approx 1$ ({\it solid curves}) and for
$\bar{x}_i = 0.5 $ ({\it dashed curves}).  The curves in the top two
panels are calculated in the same volume and with the same number of
objects, $58$, as the SDF photometric sample, and the curves in the
bottom panel are calculated from $250$ emitters in an area that is
slightly larger than the upcoming SXDS.  These curves
assume that $m \, \exp(-\tau_{\alpha}(\nu_0)) > 5\times10^{10}~\Msun$,
and there are $8$ curves in each set.  The thick solid line with
circles in the top two panels is the 2-D correlation of the SDF
photometric sample at $z = 6.6$ \citep{kashikawa06}.  The insets
magnify the region $4 \; \Mpc/h < r < 20 \; \Mpc/h$.
\label{fig:boot}}
\end{figure}

\citet{kashikawa06} measured the angular correlation function $w_{\rm
E}(r)$ from the SDF photometric sample of $58$ emitters.
Interestingly, they find that $w_{\rm E}(r)$ is consistent with no
clustering ({see the connected circles in Fig. \ref{fig:boot} for
this correlation function}).  This result may allow us to put
constraints on $\bar{x}_i$ at $z = 6.6$.  In addition, the $z =
6.6$ LAE sample will increase in size by a factor of five in the coming
year with the Subaru/XMM-Newton Deep Survey (SXDS) \citep{ouchi05}.

To proceed, we estimate the mean value of $w_{\rm E}(r)$ as well as
the covariance in $w_{\rm E}(r)$ between radial bins, $\bfC_{\rm E}$,
from many mock surveys. Given the dimensions of the survey, we
generate these mock catalogs in as many spatially independent volumes
as our $186~\Mpc$ simulation box allows, and we do this computation
for $\tau_{\alpha}(\nu_0)$ calculated along the $\hat{i}$, $\hat{j}$,
and $\hat{k}$ directions. (For the SDF, $20$
spatially independent surveys can fit into the simulation box, leading
to $60$ mock catalogs.  In practice, we construct many times more
overlapping mock catalogs than this number to obtain all the
information that is available from the simulation box.)  In addition,
we compute $w_{\rm E}(r)$ and $\bfC_{E}$ from all the halos above
$m_{\rm min}$ in each mock survey region, and then we
subtract out the shot noise contribution to determine the cosmological
part of the covariance matrix.  This procedure takes advantage of the
fact that the cosmological contribution to $w_{\rm E}(r)$ and
$\bfC_{E}$ does not depend on $f_{\rm E}$, allowing us to reduce the
uncertainty in our estimates for these quantities.

Figure \ref{fig:wcorr} plots predictions for the correlation function
of LAEs at $z = 6.6$ and $m_{\rm min} = 7\times10^{10}~\Msun$.  The
pessimistic estimate in \citet{kashikawa06} for the contamination
fraction of the SDF photometric sample is $F_c = 0.27$, and
\citet{kashikawa06} estimates that the contamination is probably
closer to $F_c = 0.16$.  We set $F_c = 0$ for the spectroscopic survey
curves ({\it bottom panel}) and $F_c = 0.25$ for the photometric
surveys ({\it middle and top panels}).  We also lower the number of
emitters in the mock spectroscopic surveys by $1 - F_c$, where $F_c =
0.25$ is the contamination fraction in the mock photometric sample.
Foreground contamination will bias the estimate for the measurement of
the correlation function by the factor $(1 -F_c)^2$.  Rather than plot
biased curves for $w_{\rm E}$ in Figure \ref{fig:wcorr}, we instead
divide the Poisson errors by the appropriate factor to account for
contamination.

The top panel in Figure \ref{fig:wcorr} displays the average
correlation function for several clustering models, generated in mock
surveys with the same dimensions as the SDF ($34'\times 27'\times 130~
\AA$) and with $58$ emitters -- the number of LAEs in the SDF
photometric sample.  The thick error bars in Figure \ref{fig:wcorr}
account for shot noise and the thin error bars include both shot noise
and cosmic variance.  Note that the cosmic variance errors in the top
panel are important, particularly on large scales.  The impact of
cosmic variance is relatively independent of the flux sensitivity of
the survey.  Therefore, a larger survey volume than the SDF is
necessary to mitigate its effect.

However, Figure \ref{fig:wcorr} suggests that current observations in
the SDF can distinguish a model with $\bar{x}_i = 0.5$ from one with
$\bar{x}_i \approx 1$.  Figure \ref{fig:boot} illustrates more
explicitly the ability to constrain reionization with current and
upcoming surveys.  Note that the y-axis in Figure \ref{fig:boot} is
linear rather than log as in Figure \ref{fig:wcorr}.  The top two
panels in Figure \ref{fig:boot} show $w_{\rm E}$ measured from
different locations in our box for similar survey specifications as in
Figure \ref{fig:wcorr} and with $F_c = 0$ and $F_c = 0.25$.  The
dashed curves are $8$ randomly selected sets of $w_{\rm E}$ for
$\bar{x}_i = 0.5$, and the solid ones are the same for $\bar{x}_i \approx 1$.
Both groups of curves in the top two panels in Figure \ref{fig:boot}
are measured from independent volumes in the simulation box, and each
curve is calculated from $58$ LAEs. The two sets of curves are fairly
distinct in the top and middle panels in Figure \ref{fig:boot},
suggesting that $\bar{x}_i < 0.5$ can be distinguished from $\bar{x}_i
\approx 1$ using current data.

The middle panel in Figure \ref{fig:wcorr} shows the correlation
function for mock surveys similar to the $z = 6.6$ SXDS that will be
completed in the next year.  We compute this function in a $1.4$
sq. deg. area (the angular size of the $186$ Mpc simulation box), with
a $130\, \AA$ narrow band filter, and from $250$ emitters.  The SXDS
is instead $1$ sq. deg., but should have a similar number of emitters.
Here, the uncertainty in $w_{\rm E}$ has been reduced substantially
compared to the top panel.  Also see the bottom panel in Figure
\ref{fig:boot}.  With an SXDS-like survey, even $\bar{x}_i = 0.7$ can,
on average, be distinguished from $\bar{x}_i \approx 1$.

The bottom panel in Figure \ref{fig:wcorr} is for a luminosity-limited
spectroscopic survey in the $1.4$ sq. deg. area and with $F_c = 0$,
with the same $m_{\rm min}$ as the middle panel, and with the number
of spectroscopically confirmed emitters reduced by the factor $1 -
0.25$ from the mock photometric surveys in the middle panel.  Since it
is more expensive to perform spectroscopy, it is interesting to
estimate how much a spectroscopic sample improves the potential to
detect reionization.  The only difference between the top panel and
the middle panel is that the shot-noise errors have been reduced by
the factor $\approx (1 -F_c)$.  A spectroscopic sample is not
significantly more sensitive to $w_{\rm E}$ than a photometric sample
if $F_c = 0.25$.  The benefits of a spectroscopic survey increase for
larger contamination fractions.  In addition, spectroscopic surveys
open up a third dimension for study.  For narrow band LAE surveys, the
narrow third dimension does not provide much additional information,
but for broad band Ly$\alpha$ surveys (as will be conducted with JWST) three
dimensional clustering statistics will be the way to go.

Motivated by the appearance of large voids in the LAE field, in
Appendix \ref{voids} we investigate the ability of void statistics to
detect reionization.  We find that, provided that $F_c \approx 0$, void
statistics can distinguish models that have different $\bar{x}_i$ with
comparable significance to the correlation function.

\subsection{Subaru Data}
\label{data}
We have seen that a SDF-like data set can distinguish between
different clustering models for the $z = 6.6$ LAEs.  Let us quantify
how well the real SDF data can distinguish between these different
models.  We take the $z= 6.6$ SDF photometric sample of
$58$ emitters presented in \citet{taniguchi05} and \citet{kashikawa06}
as well as the exact survey specifications, accounting for masked
regions.  We then compute $w_{\rm E}(r)$ using the \citet{landy93}
unbiased estimator.  The connected circles in Figure
\ref{fig:boot} represent the $z = 6.6$ correlation function of
emitters presented in \citet{kashikawa06}.

We compute the likelihoods of different theoretical models for
$w_{\rm E}$ and $\bfC_{\rm E}$ given the data (and assuming Gaussian
statistics).  These models are constructed in the same way as in
Section \ref{detectability}.  We assume a survey that has the same
angular dimensions as SDF as well as a line-of-sight window function
$W(z)$ that is a Gaussian in $\nu$ with FWHM $132~\AA$ and that is centered at
$z = 6.6$ (such that $m \, \exp(-\tau_{\alpha}(\nu_0)) \, W(z_{m}) >
m_{\rm min}$ in order to be observed).  This window function is a fair
approximation to the window provided by the $9210~\AA$ narrow band
filter on Suprime-Cam, but we find that even if we use a tophat for
the window function, our conclusions are unchanged, implying that the
exact functional form of $W(z)$ is unimportant.

Let us compare the likelihoods of models in which $\bar{x}_i \approx
1$ to models in which the universe is significantly neutral.  Table
\ref{table1} summarizes our results.  We find that if $m_{\rm min} =
7\times10^{10}~\Msun$ and $F_c = 0.25$ (adjusting $f_{\rm E}$ to yield
the observed $\bar{n}_{\rm E}$), a universe with $\bar{x}_i \approx 1$
is favored by the data over a universe with $\bar{x}_i = 0.5$ at
$3.3$-$\sigma$ (with $\bar{x}_i = 0.7$ at $2.2$-$\sigma$).  For a
model with $m_{\rm min} = 3\times10^{10}~\Msun$ and $F_c = 0.25$
(which requires $f_{\rm E} = 0.02$ when $\bar{x}_i \approx 1$), a
universe with $\bar{x}_i \approx 1$ is favored over a universe with
$\bar{x}_i = 0.5$ at $3.0$-$\sigma$.  

\emph{Using reionization model (i), if we marginalize over $m_{\rm
min}$ and $F_c$ with the priors $m_{\rm min} > 3\times10^{10}~\Msun$
and $F_c > 0.25$ (with $v_{\rm w} = 0$), the maximum likelihood model with
$\bar{x}_i \approx 1$ is favored over the maximum likelihood model
with $\bar{x}_i < 0.5$ at $3.0$-$\sigma$ ($\bar{x}_i < 0.7$ at
$1.9$-$\sigma$).}  If we relax the prior to $m_{\rm min} >
1\times10^{10}~\Msun$ (such that at equality $f_{\rm E} =
4\times10^{-3}$) than $\bar{x}_i < 0.5$ is then disfavored at
$2$-$\sigma$.  If we used model (ii) [model (iii)] to do the
calculation discussed in the first sentence in this paragraph, a model
with $\bar{x}_i \approx 1$ is preferred over a model with $\bar{x}_i <
0.5$ at $2.9$-$\sigma$ [$2.8$-$\sigma$].  Also, for reionization model
(i) and the toy galactic wind model in which all LAEs have $v_{\rm w}
= 400$ km s$^{-1}$, $\bar{x}_i < 0.5$ is disfavored at $2.1$-$\sigma$.

It is useful to note that the clustering data of $z = 6.6$ LAEs can
even be used to constrain the intrinsic clustering of emitters if we
assume that $\bar{x}_i \approx 1$ at $z = 6.6$.  For example, the current data
favors $m_{\rm min} = 3\times10^{10}~\Msun$ over $m_{\rm min} =
7\times10^{10}~\Msun$ at $1.9$-$\sigma$ assuming $F_c = 0.25$.  The
current data favors $F_c = 0.5$ over $F_c = 0.25$ at $1.5$-$\sigma$
for $m_{\rm min} = 3\times10^{10}~\Msun$.

\begin{table}
\caption{$\chi^2 \equiv -2 \;\log{\cal L}$ for different clustering models given the
  clustering in the $z = 6.6$ SDF photometric sample.  These calculations assume model (i) for reionization.  The model marked ``wind'' assumes $v_{\rm w} = 400$ km s$^{-1}$.  Note that to
  achieve the observed abundance at $z = 6.6$ for $\bar{x}_i \approx
  1$ then, on average, $f_{\rm E} = 0.02,\; 0.1$ and $0.2$ for $m_{\rm
  min} = 3\times 10^{10}\; \Msun, \; 7\times 10^{10} \; \Msun,$ and
  $1\times 10^{11}~\Msun$, respectively.}
\begin{center}
\begin{tabular}{l c c c c}
\hline
$m_{\rm min}$ ($\Msun$)  & $\bar{x}_i$ & $\chi^2$ ($F_c = 0$) & $\chi^2$ ($F_c = \frac{1}{4}$) \\
\hline
$3\times 10^{10}$ & $1.0$ & $4.9$ & ${\bf 1.1 }$\\
& $0.7$ & $10.3$ & $4.9$\\
& $0.5$ & $16.6$ & $9.9$\\
& $0.3$ & $22.4$ & $15.1$\\
$7\times 10^{10}$ & $1.0$ & $10.5$ & $4.9$\\
& $0.7$ & $16.0$ & $9.6$\\
& $0.5$ & $23.6$ & $15.9$\\
& $0.3$ & $32.6$ & $23.9$\\
$1\times 10^{11}$ & $1.0$ & $14.2$ & $7.8$\\
& $0.7$ & $19.4$ & $12.2$\\
& $0.5$ & $29.3$ & $20.7$\\
$3\times 10^{10}$ (wind) & $0.7$ & $8.2$ & $3.2$\\
& $0.5$ & $11.5$ & $5.4$\\
& $0.3$ & $13.8$ & $7.5$\\
\hline
\end{tabular}
\end{center}
 \label{table1}
\end{table}

In this analysis, we have assumed the statistics are Gaussian.  We use
$8$ logarithmically-spaced radial bins that run between $0.4$ and $70$
Mpc to compute $w_{\rm E}$ and $\bfC_{\rm E}$.  The bins at the
smallest radii are the least Gaussian because they contain the fewest
pairs of LAEs.  If we discard the first two radial bins ($r < 1.5\;
\Mpc$), the maximum likelihood model for $\bar{x}_i \approx 1$ can be
distinguished by the SDF data from the maximum likelihood model for
$\bar{x}_i < 0.5$ at $2.6$-$\sigma$ rather than at $3.0$-$\sigma$
assuming model (i) for reionization and the priors $m_{\rm min} >
3\times10^{10}~\Msun$, $F_c > 0.25$, and $v_{\rm w} = 0$.

We ignored the effect of survey incompleteness in the above analysis.
The SDF survey is complete at $\approx 50\%$ level above the
luminosity threshold $L_{\rm min}$ (and at the $\approx 75\%$ level
above $2.5 \, L_{\rm min}$) \citep{kashikawa06}.  We have investigated
the importance of this effect by randomly discarding half of the
objects that fall below $1.5 \;L_{\rm min}$ and one fourth of objects
between $1.5 \;L_{\rm min} < L_{\rm obs, E} < 2.5 \;L_{\rm min}$, and
we find that this operation changes the cosmological part of $P_{\rm
E}$ by $<25\%$ on all scales and does not significantly alter the
conclusions in this section.  We find that incompleteness is
degenerate with the parameter $f_{\rm E}$.

In this analysis, we have also assumed that the shape of the
luminosity function was set by the shape of the mass function.  The $z
= 5.7$ and $z = 6.6$ luminosity functions are poorly constrained at
the faint end, providing good fits for faint end power-law indexes of
$\beta = 2,~ 2.5$, and $3$ \citep{shimasaku06, kashikawa06}.  The
effect of reionization on the LAEs depends on the slope of the
luminosity function near $L_{\rm min}$ because the emitters that fall
near the detection threshold are the easiest to obscure with neutral
regions.  The luminosity function in our models scales as
$\beta = 2.3$ at $m_{\rm min} = 1 \times10^{10} ~\Msun$, $\beta = 2.6$
at $m _{\rm min}= 5 \times10^{10} ~\Msun$, and $\beta = 2.8$ at
$m_{\rm min} = 1 \times10^{11} ~\Msun$.  Therefore, the range of
$m_{\rm min}$ that we have considered spans much of the relevant parameter
space for $\beta$.

The calculations in this section were aimed at understanding the
sensitivity to reionization of widefield surveys at $z =6.6$ that have
already been completed or will be finished this coming year.  These
surveys constitute a small fraction of the total Subaru observing time
in a year.  However, even with these surveys, constraints can be
placed on reionization.  The sensitivity to reionization of a mission
dedicated specifically to this science would be vastly superior.  We
discuss the prospects for a few upcoming LAE surveys in Section
\ref{surveys}.

\section{Future LAE Surveys}
\label{surveys}

We have shown that if the Universe is largely neutral at $z \lesssim
7$, observations like those being done on Subaru should be able to
study the modulation from the HII regions.  What is the prospect for
upcoming surveys to detect reionization at $z >7$?  To answer this
question, we concentrate on three instruments: (1) the Dark Ages z
Ly$\alpha$ Explorer (DAzLE), which has started observing on the Very
Large Telescope (VLT), (2) the configurable
multi-slit spectrograph MOSFIRE ({\it the Multi-Object Spectrograph
For Infra-Red Exploration}) to be commissioned on the Keck telescope
in 2009, and (3) NIRSpec on JWST.  These represent
some of the most promising instruments to target high-redshift LAEs.

Upcoming instruments take different approaches to identify
high-redshift LAEs. DAzLE uses two $10 \, \AA$, overlapping narrow
band filters in a $6'\times 6'$ FoV.  Differencing these filters makes
it possible to subtract out continuum sources, leaving just the
broad-line sources.  The effective volume for DAzLE in a single
pointing is $1600~\Mpc^3$, or $\approx 1\%$ of the volume of the SDF
observations.  However, DAzLE's flux sensitivity is below that of
other narrow band surveys because of its extremely narrow filters.
MOSFIRE uses the magnification bias of cluster lensing caustics to
improve the sensitivity to high-redshift LAEs by as much as a factor
of $50$.  MOSFIRE is capable of simultaneously taking spectra from
$45$ slits, each $7.''3$ in length.  For an $80$ hr observation of $8$
clusters, \citet{stark07b} estimate that these surveys could cover an
effective volume of $80$ Mpc$^{3}$ distributed between $z = 7.0-8.3$
-- much smaller than the volume of narrow band samples.  However,
whereas DAzLE will be sensitive to $L_{\rm obs, E} \approx 10^{42}\,
{\rm erg}\, s^{-1}$ at $z = 8$, MOSFIRE can measure line fluxes of
lensed galaxies which have $L_{\rm obs, E}\approx 5 \times10^{40}\,
{\rm erg} \, s^{-1}$ in a $10$ hr exposure.

With the launch of JWST in 2013, the NIRSpec instrument (capable of
simultaneously taking spectra of $100$ objects in a $3'\times 3'$ FoV)
will be sensitive to Ly$\alpha$ luminosities that are over an order of
magnitude smaller than terrestrial narrow band surveys and comparable
to the effective sensitivity of a MOSFIRE caustic survey.  JWST will
use the imaging camera NIRCam to select high-redshift galaxies with
the Lyman break technique, and follow up with NIRSpec to select LAEs.
This selection strategy will facilitate cross correlation studies to
isolate the bubble-induced fluctuations ({as discussed in Section
\ref{clustering}}) and allow 3-D tomography of LAEs.  See Figure
\ref{fig:maps2} for a mock survey of LAEs that has a JWST-like
sensitivity.  However, a survey with JWST that has the same area as
the panels in Figure \ref{fig:maps2} would in fact require over $100$
different pointings of both NIRCam and NIRSpec to cover this $1000$
sq. arc-min. field.  ({The box in the lower left-hand panel of
Fig. \ref{fig:maps2} is the FoV for NIRSpec.})  The current
high-redshift program for JWST outlines a deep survey in a much
smaller area, a tens of sq. arc-min. region \citep{gardner06}.

In general, it is difficult to make predictions for the statistical
significance with which upcoming instruments can detect reionization
at $z > 7$ owing to all of the uncertainties in the source properties
and their evolution.  As we have seen, there is substantial
uncertainty in the $z = 6.6$ luminosity function.  The extrapolation
of this luminosity function to higher redshifts depends sensitively
on, for example, the typical halo mass of LAEs.  If $f_{\rm E} = 1$ in
current LAE surveys (such that the observed LAEs are in the most
massive halos), then the luminosity function will evolve more quickly
with redshift than if $f_{\rm E} < 1$.  Unfortunately, the parameter
$f_{\rm E}$ is not constrained to even an order of magnitude by the $z
= 6.6$ data \citep{dijkstra06a, stark07b}.  Other factors could also
shape the evolution of the Ly$\alpha$ luminosity function such as the
build-up of dust or an evolving abundance of metal-poor stars within
galaxies.

Despite these challenges, several studies have made estimates for
$\bar{n}_{\rm E}$ as a function of redshift \citep{stark07b, barton04,
thommes05}.  For DAzLE, \citet{thommes05} and \citet{stark07b} find
that roughly $1$ LAE should be observed at $z = 8$ in a single $10$ hr
pointing of the camera.  To estimate the abundance of LAEs that will
be observed with MOSFIRE on Keck and NIRSpec on JWST is more difficult
than for DAzLE because these instruments are at least an order of
magnitude more sensitive, probing a region of the luminosity function
that has yet to be explored.  Fortunately, an estimate for the
observed number of emitters is less important for understanding the
scientific impact of MOSFIRE and JWST because the limiting factor will
be cosmic variance and not shot noise.

\begin{figure}
  \rotatebox{-90}{\epsfig{file=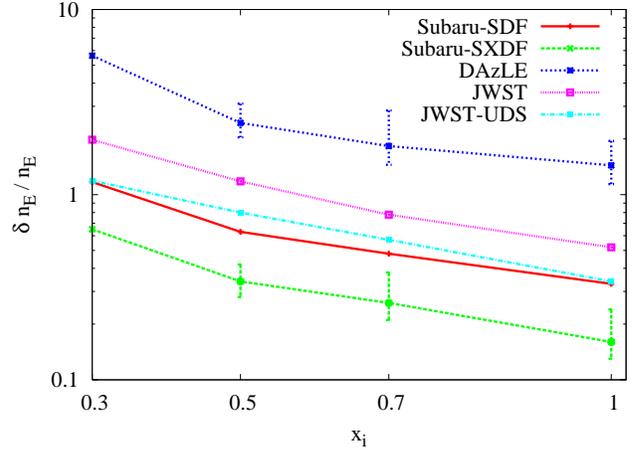, height=8.8cm}}
\caption{Cosmic fluctuations in the measured LAE number density as a
function of $\bar{x}_i$ for current and upcoming surveys (the variance
owing to shot noise has been subtracted).  Most curves assume one
pointing of the relevant imaging camera/spectrograph (SDF is one
pointing of Suprime-Cam), with the exceptions being those for the SXDF
and UDS surveys, in which the variance is calculated for the survey
volume.  The DAzLE and Subaru curves are computed assuming that
$m_{\rm min} = 5\times10^{10} ~\Msun$, and those for JWST are
calculated assuming that $m_{\rm min} = 1\times10^{10} \; \Msun$.  The
error bars on the DAzLE and SXDS curves represent the change in
$\delta n_e/n_e$ if $m_{\rm min}$ is reduced or increased by a factor
of $2$.
\label{fig:variance}}
\end{figure}

Here, we make predictions for the cosmic variance in the measured LAE
number density for upcoming surveys.  We do not perform this calculation
for a MOSFIRE cluster caustic survey owing to the complicated geometry
(see \citealt{stark07b}).  In the context of our simple model for
the LAEs, cosmic variance depends on just the parameter $m_{\rm min}$.
When fluctuations in the measured number density of emitters within a
survey volume are on the order of unity about the average abundance
$\bar{n}_{\rm E}$, it will be difficult to understand reionization
without a larger survey volume.  For example, these fluctuations will
prevent a precise measurement of the luminosity function. 

 The curves in Figure \ref{fig:variance} represent the standard
deviation in the number density of LAEs (with the shot noise component
subtracted out) measured with current and upcoming surveys (and
assuming no foreground contamination).  Note that neutral gas
increases the standard deviation significantly.  The DAzLE and Subaru
curves are computed assuming that $m_{\rm min} = 5\times10^{10}
~\Msun$ (which is a conservative choice and requires $f_{\rm E} <
0.1$), and the ones for JWST are calculated assuming $m_{\rm min} =
1\times10^{10}~ \Msun$.  All of these curves are determined by first
convolving the LAE maps with a window function that has the same
volume specifications as these surveys/instruments and, then, by
computing the variance of the windowed field.  For JWST, which will
take spectra over a broad range of wavelengths with NIRSpec, we have
assumed that the line-of-sight width used to estimate $\bar{n}_{\rm
E}(z)$ is $100$ Mpc ($\Delta z \sim 0.3$).  The choice of $100$ Mpc is
motivated by the short duration of reionization in simulations of reionization
($0.5 < \bar{x}_i \lesssim 1$ spans $\Delta z \approx 1$) and the
desire to study LAEs for a few different ionized fractions during
reionization.

The curves in Figure \ref{fig:variance} are computed from snapshots
between $z= 7$ and $z= 8$. The value of $\delta \bar{n}_{\rm
E}/\bar{n}_{\rm E}$ depends modestly on the redshift of the survey.
For example, for the SXDS, for $m_{\rm min} = 5\times10^{10} ~\Msun$
and $\bar{x}_i \approx 1$ then $\delta n_{\rm E}/n_{\rm E} =
0.13$ at $z = 7$ whereas $\delta n_{\rm E}/n_{\rm E} = 0.17$ at $z =
8$.\footnote{Note that the SXDS curve in Figure \ref{fig:variance}
assumes a square survey in a $1$ sq. deg. FoV.  SXDS is instead a
cross-shaped survey with the same volume.}  In addition, if we vary the
value of $m_{\rm min}$, the results do not change significantly.  The
error bars on the DAzLE and Subaru-SXDF curves in Figure
\ref{fig:variance} represent the change in $\delta n_e/n_e$ if the
fiducial $m_{\rm min}$ is reduced or increased by a factor of two.

The volume surveyed by DAzLE is much smaller than the volume of a
$20\; \Mpc$ bubble.  For some pointings, the surveyed volume may
consist entirely of a neutral region (and zero emitters will be
observed), and, for other pointings, it may fall entirely within an
HII region.  Figure \ref{fig:variance} shows that in order to
constrain the cosmic fluctuations in the luminosity function to
$50\%$, DAzLE needs $N_p \approx 10$ non-contiguous pointings when
$\bar{x}_i \approx 1$ and significantly more pointings when the
Universe is largely neutral.  (Note that $(\delta n_{\rm E})_{N_p}
\approx \delta n_{\rm E}/\sqrt{N_p}$ for non-contiguous pointings.)
Of course, this observing strategy is optimized for reducing the
cosmic variance.  Shot noise may be as important for DAzLE.

With JWST, one might hope to be able to place a tight constraint on
the LAE luminosity function.  However, for a $20\%$ constraint, JWST
needs $N_p \approx 10$ non-contiguous pointings when $\bar{x}_i
\approx 1$ (and $N_p \approx 50$ when $\bar{x}_i \approx 0.5$).

Much of the high-redshift data for JWST will be gathered in the
Ultra-Deep Survey (UDS), which will be a contiguous field spanning
tens of sq. arc-min. \citep{gardner06}.  Here, we generously assume a
square  field of $80$ sq. arc-min ($\approx 9$ tiles observed with both
NIRCam and NIRSpec).  Figure \ref{fig:variance} shows that this
observation is far from optimal for constraining the luminosity
function with $\delta n_{\rm E}/n_{\rm E} \gtrsim 0.3$.
A different observing strategy is necessary for JWST to derive tight
constraints on reionization with LAEs. 

The largest volume LAE survey currently being conducted at $z > 8$ is
the MOIRCS Deep Survey on Subaru, which targets LAEs at $z = 8.8$.
While not as sensitive as DAzLE, its volume of $7'\times4'\times 100
\;\AA$ is an order of magnitude larger than that of DAzLE
\citep{ouchi_proc}.  If we had plotted a curve of $\delta n_{\rm
E}/n_{\rm E}$ for MOIRCS in Figure \ref{fig:variance}, this curve
would fall between the curves for the Subaru-SDF and for DAzLE.

In this section, we have seen that cosmic variance will be a
significant concern for upcoming LAE surveys at $z > 7$.  Degree-scale
surveys make the cosmic variance manageable for this science.

\section{Conclusions}

LAE surveys are probing increasingly higher redshifts.  Not only will
these surveys inform us about high-redshift galaxies, but they have
the potential to be the first observations to unambiguously detect
patchy reionization.  The tens of Mpc HII regions during reionization
modulate the observed distribution of LAEs and boost their observed
clustering.  We have shown that this effect on the angular correlation
function (Section \ref{clustering}) or the void probability
distribution (Appendix \ref{voids}) can be well in excess of the
intrinsic clustering of halos in the concordance cosmology.  This enhanced
clustering depends most strongly on $\bar{x}_i$, but also somewhat on
the morphology of the HII regions during reionization.  Observing
enhanced clustering would confirm the prediction that the HII regions
during reionization are large.

Even the current $z = 6.6$ Subaru LAE survey, which has a photometric
sample of only $58$ LAEs in a $0.25$ deg$^2$ field, can place
constraints on the reionization process.  We find that the angular
correlation function of the SDF photometric sample of $z = 6.6$ LAEs
favors an ionized universe over a universe with $\bar{x}_i < 0.5$ at a
$3$-$\sigma$ confidence level ($2$-$\sigma$ if all emitters have
strong galactic winds). This constraint is both competitive with other
constraints on $\bar{x}_i$ obtained from GRBs and from LAEs
\citep{totani05, malhotra05}, and it rules out the picture of the $z =
6$ quasars expanding into a neutral IGM.  In addition, this is the
first constraint on the ionized fraction that consistently accounts
for patchy reionization -- the favored model for how the universe is
ionized.  Observations in the next year in the SDXS will increase the
$z = 6.6$ sample by a factor of $5$ and place even stronger
constraints on $\bar{x}_i$.

Detecting reionization through LAE clustering also offers a simple
consistency check for whether the observed correlations owe to
reionization: observe the same field with a second selection technique
in addition to selecting objects by their Ly$\alpha$ emission.  If the
Ly$\alpha$-selected galaxies show enhanced clustering relative to the
galaxies selected with the other technique, the evidence for reionization
is strengthened.  Similarly, a comparison of the phases of Fourier
modes between the two galaxy samples can be used to detect reionization.

While measuring enhanced clustering is the most fool-proof method to
detect reionization with LAEs, the presence of neutral regions in the
IGM also influences the properties of the Ly$\alpha$ line and the
luminosity function.  Observing evolution that is consistent with
reionization occurring in all of these different statistics would
strengthen the argument for reionization.  We have made predictions
for the scale of these effects.  We show in Appendix \ref{lineprofile}
that the effect of reionization on the average line profile is less
significant than most previous studies have found.  This result owes
to the much larger HII regions that arise from properly treating
source clustering and because emitters that are not significantly
obscured by the neutral regions are preferentially observed.  Because
of these effects, combined with the uncertainty in the astrophysical
processes that determine the Ly$\alpha$ line shape, the consequences
of reionization will be difficult to isolate in the line profile.  

The impact of reionization on the luminosity function is a more
promising diagnostic than the line profile.  Its effect on the
luminosity function is similar among the three reionization models we
have considered, if we compare at fixed $\bar{x}_i$.  In all models
and at all $\bar{x}_i$, the LAE luminosity function is suppressed by a
factor that is fairly constant as a function of luminosity.  Our
predictions for the impact of reionization on the luminosity function
are fairly robust to our assumptions concerning the bias and intrinsic
luminosity of emitters, but it may be difficult to distinguish
evolution in the luminosity function that owes to reionization from
evolution owing to changing intrinsic properties of the emitters.

It is likely that most of reionization occurs at higher redshifts than
probed by current samples.  Several LAE surveys will target $z > 7$
LAEs in the coming years.  Unfortunately, these upcoming studies are
not optimal for measuring the LAE luminosity function or for detecting
reionization-induced clustering, having fields of view that are too
small to measure this effect precisely.  Even the multi-billion dollar
satellite JWST has this design flaw.  JWST's enhanced sensitivity over
that of current telescopes still makes it useful for this science.
However, a multi-billion dollar space mission is unnecessary to
understand reionization with LAEs.  Widefield observations like those
on Subaru, but that target $z >7$, will be able to put constraints on
the reionization process.

\section{acknowledgments} 
We thank Nobunari Kashikawa for answering our questions about the $z =
6.6$ SDF LAE survey, Adam Lidz for many helpful conversations, Masami
Ouchi for useful discussions about the Subaru LAE surveys and helpful
comments, Volker Springel for providing his Lean Gadget--2 code,
Alexander Tchekhovskoy for answering MMs numerous computing questions,
and Oliver Zahn for helpful comments and for providing the 2LPT
displacement fields.  We would also like to thank Steve Furlanetto for
pointing out a mistake in a previous version of this manuscript.  MM
acknowledges support through an NSF graduate student fellowship.  The
authors are also supported by the David and Lucile Packard Foundation,
the Alfred P. Sloan Foundation, and grants AST-0506556 and
NNG05GJ40G.

\bibliographystyle{mn2e}

\begin{thebibliography}{}

\bibitem[\protect\citeauthoryear{Babich \& Loeb}{Babich \&
  Loeb}{2006}]{babich06}
Babich D.,  Loeb A.,  2006, \apj, 640, 1

\bibitem[\protect\citeauthoryear{Barton et~al.,}{Barton
  et~al.}{2004}]{barton04}
Barton E.~J.,  et~al., 2004, \apj, 604, L1

\bibitem[\protect\citeauthoryear{{Becker}, {Rauch} \& {Sargent}}{{Becker}
  et~al.}{2006}]{becker06}
{Becker} G.~D.,  {Rauch} M.,    {Sargent} W.~L.~W.,  2006, astro-ph/0607633

\bibitem[\protect\citeauthoryear{{Becker} et~al.,}{{Becker}
  et~al.}{2001}]{becker01}
{Becker} R.~H.,  et~al., 2001, \aj, 122, 2850

\bibitem[\protect\citeauthoryear{{Bolton} \& {Haehnelt}}{{Bolton} \&
  {Haehnelt}}{2007}]{bolton06}
{Bolton} J.~S.,  {Haehnelt} M.~G.,  2007, \mnras, 374, 493

\bibitem[\protect\citeauthoryear{{Casali} et~al.,}{{Casali}
  et~al.}{2006}]{casali06}
{Casali} M.,  et~al., 2006, in Ground-based and Airborne Instrumentation for
  Astronomy. Edited by McLean, Ian S.; Iye, Masanori. Proceedings of the SPIE,
  Volume 6269, pp. (2006). {HAWK-I: the new wide-field IR imager for the VLT}

\bibitem[\protect\citeauthoryear{{Croton} et~al.,}{{Croton}
  et~al.}{2004}]{croton04}
{Croton} D.~J.,  et~al., 2004, \mnras, 352, 828

\bibitem[\protect\citeauthoryear{{Cuby}, {Hibon}, {Lidman}, {Le F{\`e}vre},
  {Gilmozzi}, {Moorwood} \& {van der Werf}}{{Cuby} et~al.}{2007}]{cuby06}
{Cuby} J.-G.,  {Hibon} P.,  {Lidman} C.,  {Le F{\`e}vre} O.,  {Gilmozzi} R.,
  {Moorwood} A.,    {van der Werf} P.,  2007, \aap, 461, 911

\bibitem[\protect\citeauthoryear{{de Vos}}{{de Vos}}{2004}]{deVos04}
{de Vos} M.,  2004, http://www.lofar.org/PDF/LOFAR-P1-Baseline2.0.pdf

\bibitem[\protect\citeauthoryear{{Dekel} \& {Woo}}{{Dekel} \&
  {Woo}}{2003}]{dekel03}
{Dekel} A.,  {Woo} J.,  2003, \mnras, 344, 1131

\bibitem[\protect\citeauthoryear{{Dijkstra}, {Haiman} \& {Spaans}}{{Dijkstra}
  et~al.}{2006}]{dijkstra06a}
{Dijkstra} M.,  {Haiman} Z.,    {Spaans} M.,  2006, \apj, 649, 14

\bibitem[\protect\citeauthoryear{{Dijkstra}, {Lidz} \& {Wyithe}}{{Dijkstra}
  et~al.}{2007}]{dijkstra06b}
{Dijkstra} M.,  {Lidz} A.,    {Wyithe} S.,  2007, Submitted to MNRAS
  (astro-ph/0701667)

\bibitem[\protect\citeauthoryear{{Fan} et~al.,}{{Fan}  et~al.}{2006}]{fan06}
{Fan} X.,  et~al., 2006, \aj, 132, 117

\bibitem[\protect\citeauthoryear{{Furlanetto}, {McQuinn} \&
  {Hernquist}}{{Furlanetto} et~al.}{2005}]{furl-models}
{Furlanetto} S.~R.,  {McQuinn} M.,    {Hernquist} L.,  2005, \mnras, pp 1043--+

\bibitem[\protect\citeauthoryear{{Furlanetto}, {Oh} \& {Briggs}}{{Furlanetto}
  et~al.}{2006}]{furl-rev}
{Furlanetto} S.~R.,  {Oh} S.~P.,    {Briggs} F.~H.,  2006, Physics Reports,
  433, 181

\bibitem[\protect\citeauthoryear{{Furlanetto}, {Zaldarriaga} \&
  {Hernquist}}{{Furlanetto} et~al.}{2004a}]{furlanetto04a}
{Furlanetto} S.~R.,  {Zaldarriaga} M.,    {Hernquist} L.,  2004a, \apj, 613, 1

\bibitem[\protect\citeauthoryear{{Furlanetto}, {Hernquist} \&
  {Zaldarriaga}}{{Furlanetto} et~al.}{2004c}]{furlanetto04c}
{Furlanetto} S.~R.,  {Hernquist} L.,    {Zaldarriaga} M.,  2004c, \mnras, 354,
  695

\bibitem[\protect\citeauthoryear{{Furlanetto}, {Sokasian} \&
  {Hernquist}}{{Furlanetto} et~al.}{2004b}]{furlanetto04b}
{Furlanetto} S.~R., {Sokasian} A.,  {Hernquist} L.,  2004b, \mnras, 347,
  187

\bibitem[\protect\citeauthoryear{Furlanetto, Zaldarriaga \&
  Hernquist}{Furlanetto et~al.}{2006}]{furl-galaxies05}
Furlanetto S.~R.,  Zaldarriaga M.,    Hernquist L.,  2006, \mnras, 365, 1012

\bibitem[\protect\citeauthoryear{{Gardner} et~al.,}{{Gardner}
  et~al.}{2006}]{gardner06}
{Gardner} J.~P.,  et~al., 2006, Space Science Reviews, 123, 485

\bibitem[\protect\citeauthoryear{Haiman}{Haiman}{2002}]{haiman02}
Haiman Z.,  2002, \apj, 576, L1

\bibitem[\protect\citeauthoryear{Haiman \& Cen}{Haiman \& Cen}{2005}]{haiman04}
Haiman Z.,  Cen R.,  2005, \apj, 623, 627

\bibitem[\protect\citeauthoryear{{Hansen} \& {Oh}}{{Hansen} \&
  {Oh}}{2006}]{hansen06}
{Hansen} M.,  {Oh} S.~P.,  2006, \mnras, 367, 979

\bibitem[\protect\citeauthoryear{Horton, Parry, Bland-Hawthorn, Cianci, King,
  McMahon \& Medlen}{Horton et~al.}{2004}]{horton04}
Horton A.,  Parry I.,  Bland-Hawthorn J.,  Cianci S.,  King D.,  McMahon R.,
  Medlen S.,  2004, PROC.SPIE INT.SOC.OPT.ENG., 5492, 1022

\bibitem[\protect\citeauthoryear{{Hui}, {Haiman}, {Zaldarriaga} \&
  {Alexander}}{{Hui} et~al.}{2002}]{hui02}
{Hui} L.,  {Haiman} Z.,  {Zaldarriaga} M.,    {Alexander} T.,  2002, \apj, 564,
  525

\bibitem[\protect\citeauthoryear{{Iliev}, {Shapiro} \& {Raga}}{{Iliev}
  et~al.}{2005}]{iliev-mh}
{Iliev} I.~T.,  {Shapiro} P.~R.,    {Raga} A.~C.,  2005, \mnras, 361, 405

\bibitem[\protect\citeauthoryear{{Iye} et~al.,}{{Iye} et~al.}{2007}]{iye06}
{Iye} M.,  et~al.,  2007, Nature, 443, 186

\bibitem[\protect\citeauthoryear{{Kashikawa} et~al.,}{{Kashikawa}
  et~al.}{2006}]{kashikawa06}
{Kashikawa} N.,  et~al., 2006, \apj, 648, 7

\bibitem[\protect\citeauthoryear{{Kauffmann} et~al.,}{{Kauffmann}
  et~al.}{2003}]{kauffmann03}
{Kauffmann} G.,  et~al., 2003, \mnras, 341, 33

\bibitem[\protect\citeauthoryear{Keating \& Miller}{Keating \&
  Miller}{2006}]{keating05}
Keating B.,  Miller N.,  2006, New Astron. Rev., 50, 184

\bibitem[\protect\citeauthoryear{{Kodaira} et~al.,}{{Kodaira}
  et~al.}{2003}]{kodaira03}
{Kodaira} K.,  et~al., 2003, \pasj, 55, L17

\bibitem[\protect\citeauthoryear{{Landy} \& {Szalay}}{{Landy} \&
  {Szalay}}{1993}]{landy93}
{Landy} S.~D.,  {Szalay} A.~S.,  1993, \apj, 412, 64

\bibitem[\protect\citeauthoryear{{Lidz}, {McQuinn}, {Zaldarriaga}, {Hernquist}
  \& {Dutta}}{{Lidz} et~al.}{2007}]{lidz07}
{Lidz} A.,  {McQuinn} M.,  {Zaldarriaga} M.,  {Hernquist} L.,    {Dutta} S.,
  2007, astro-ph/0703667

\bibitem[\protect\citeauthoryear{Lidz, Oh \& Furlanetto}{Lidz
  et~al.}{2006}]{lidz06a}
Lidz A.,  Oh S.~P.,    Furlanetto S.~R.,  2006, \apj, 639, L47

\bibitem[\protect\citeauthoryear{Liu, Bi, Feng \& Fang}{Liu
  et~al.}{2006}]{lui06}
Liu J.,  Bi H.,  Feng L.-L.,    Fang L.-Z.,  2006, \apj, 645, L1

\bibitem[\protect\citeauthoryear{{Loeb} \& {Rybicki}}{{Loeb} \&
  {Rybicki}}{1999}]{loeb99}
{Loeb} A.,  {Rybicki} G.~B.,  1999, \apj, 524, 527

\bibitem[\protect\citeauthoryear{{Madau} \& {Rees}}{{Madau} \&
  {Rees}}{2000}]{madau00}
{Madau} P.,  {Rees} M.~J.,  2000, \apjl, 542, L69

\bibitem[\protect\citeauthoryear{{Malhotra} \& {Rhoads}}{{Malhotra} \&
  {Rhoads}}{2004}]{malhotra04}
{Malhotra} S.,  {Rhoads} J.~E.,  2004, \apjl

\bibitem[\protect\citeauthoryear{{Malhotra} \& {Rhoads}}{{Malhotra} \&
  {Rhoads}}{2006a}]{malhotra06}
{Malhotra} S.,  {Rhoads} J.~E.,  2006a, \apjl, 647, L95

\bibitem[\protect\citeauthoryear{{Malhotra} \& {Rhoads}}{{Malhotra} \&
  {Rhoads}}{2006b}]{malhotra05}
{Malhotra} S.,  {Rhoads} J.~E.,  2006b, \apjl, 647, L95

\bibitem[\protect\citeauthoryear{{McPherson} et~al.,}{{McPherson}
  et~al.}{2006}]{mcpherson06}
{McPherson} A.~M.,  et~al., 2006, in Ground-based and Airborne Telescopes.
  Edited by Stepp, Larry M.. Proceedings of the SPIE, Volume 6267, pp. (2006).
  {VISTA: project status}

\bibitem[\protect\citeauthoryear{{McQuinn}, {Furlanetto}, {Hernquist}, {Zahn}
  \& {Zaldarriaga}}{{McQuinn} et~al.}{2005}]{mcquinn05}
{McQuinn} M.,  {Furlanetto} S.~R.,  {Hernquist} L.,  {Zahn} O.,
  {Zaldarriaga} M.,  2005, \apj, 630, 643

\bibitem[\protect\citeauthoryear{{McQuinn}, {Lidz}, {Zahn}, {Dutta},
  {Hernquist} \& {Zaldarriaga}}{{McQuinn} et~al.}{2006}]{mcquinn06b}
{McQuinn} M.,  {Lidz} A.,  {Zahn} O.,  {Dutta} S.,  {Hernquist} L.,
  {Zaldarriaga} M.,  2006, Submitted to MNRAS (astro-ph/0610094)

\bibitem[\protect\citeauthoryear{{McQuinn}, {Zahn}, {Zaldarriaga}, {Hernquist}
  \& {Furlanetto}}{{McQuinn} et~al.}{2006}]{mcquinn06}
{McQuinn} M.,  {Zahn} O.,  {Zaldarriaga} M.,  {Hernquist} L.,    {Furlanetto}
  S.~R.,  2006, \apj, 653, 815

\bibitem[Mesinger \& Haiman(2004)]{mesinger04} Mesinger, A., \& 
Haiman, Z.\ 2004, \apjl, 611, L69 

\bibitem[Mesinger et al.(2007)]{mesinger07} Mesinger, A., \& 
Haiman, Z.\ 2007, \apj, 660, 923 

\bibitem[\protect\citeauthoryear{{Miralda-Escude}}{{Miralda-Escude}}{1998}]{mi%
ralda98}
{Miralda-Escude} J.,  1998, \apj, 501, 15

\bibitem[\protect\citeauthoryear{{Morales} \& {Hewitt}}{{Morales} \&
  {Hewitt}}{2004}]{morales03}
{Morales} M.~F.,  {Hewitt} J.,  2004, \apj, 615, 7

\bibitem[Neufeld(1991)]{neufeld91} Neufeld, D.~A.\ 1991, \apjl, 
370, L85 

\bibitem[\protect\citeauthoryear{{Osterbrock}}{{Osterbrock}}{1989}]{osterbrock%
89}
{Osterbrock} D.~E.,  1989, {Astrophysics of Gaseous Nebulae and Active Galactic
  Nuclei}.
Sausalito, CA: University Science Books

\bibitem[\protect\citeauthoryear{{Ouchi et~al.,}}{Ouchi et~al.}{2005}]{ouchi05} 
{Ouchi} M., et al., 2005, \apj, 620, L1 

\bibitem[\protect\citeauthoryear{{Ouchi et~al.,}}{Ouchi et~al.}{2007}]{ouchi_proc} 
{Ouchi} M., {Tokoku} C., {Shimasaku}, K., {Ichikawa}, T. 2007, ASP Conf. Ser. in press

\bibitem[\protect\citeauthoryear{Page et~al.,}{Page  et~al.}{2006}]{page06}
Page L.,  et~al., 2006, astro-ph/0603450

\bibitem[\protect\citeauthoryear{{Partridge} \& {Peebles}}{{Partridge} \&
  {Peebles}}{1967}]{partridge67}
{Partridge} R.~B.,  {Peebles} P.~J.~E.,  1967, \apj, 147, 868

\bibitem[\protect\citeauthoryear{{Pen}, {Wu} \& {Peterson}}{{Pen}
  et~al.}{2005}]{pen04}
{Pen} U.~L.,  {Wu} X.~P.,    {Peterson} J.,  2005, Submitted to CJAA
  (astro-ph/0404083)

\bibitem[Pritchard et al.(2007)]{pritchard07} Pritchard, J.~R., 
Furlanetto, S.~R., \& Kamionkowski, M.\ 2007, \mnras, 374, 159 

\bibitem[\protect\citeauthoryear{Rhoads et~al.,}{Rhoads
  et~al.}{2004}]{rhoads04}
Rhoads J.~E.,  et~al., 2004, \apj, 611, 59

\bibitem[\protect\citeauthoryear{{Santos} et~al.,}{{Santos}
  et~al.}{2003}]{santos03}
{Santos} M.~G.,  et~al., 2003, \apj, 598, 756

\bibitem[\protect\citeauthoryear{Santos}{Santos}{2004}]{santos04}
Santos M.~R.,  2004, \mnras, 349, 1137

\bibitem[\protect\citeauthoryear{{Shapley}, {Steidel}, {Pettini} \&
  {Adelberger}}{{Shapley} et~al.}{2003}]{shapley03}
{Shapley} A.~E.,  {Steidel} C.~C.,  {Pettini} M.,    {Adelberger} K.~L.,  2003,
  \apj, 588, 65

\bibitem[{{Sheth} \& {Lemson}(1999)}]{sheth99}
{Sheth}, R.~K., \& {Lemson}, G. 1999, \mnras, 305, 946

\bibitem[\protect\citeauthoryear{{Sheth} \& {Tormen}}{{Sheth} \&
  {Tormen}}{2002}]{sheth02}
{Sheth} R.~K.,  {Tormen} G.,  2002, \mnras, 329, 61

\bibitem[Shimasaku et al.(2006)]{shimasaku06} Shimasaku, K., et 
al.\ 2006, \pasj, 58, 313 

\bibitem[\protect\citeauthoryear{{Sokasian}, {Abel}, {Hernquist} \&
  {Springel}}{{Sokasian} et~al.}{2003}]{sokasian03}
{Sokasian} A.,  {Abel} T.,  {Hernquist} L.,    {Springel} V.,  2003, \mnras,
  344, 607

\bibitem[\protect\citeauthoryear{{Sokasian}, {Abel} \& {Hernquist}}{{Sokasian}
  et~al.}{2001}]{sokasian01}
{Sokasian} A.,  {Abel} T.,    {Hernquist} L.~E.,  2001, New Astronomy, 6, 359

\bibitem[\protect\citeauthoryear{{Sokasian}, {Yoshida}, {Abel}, {Hernquist} \&
  {Springel}}{{Sokasian} et~al.}{2004}]{sokasian04}
{Sokasian} A.,  {Yoshida} N.,  {Abel} T.,  {Hernquist} L.,    {Springel} V.,
  2004, \mnras, 350, 47

\bibitem[\protect\citeauthoryear{{Spergel} et~al.,}{{Spergel}
  et~al.}{2003}]{spergel03}
{Spergel} D.~N.,  et~al., 2003, \apjs, 148, 175

\bibitem[\protect\citeauthoryear{{Spergel} et~al.,}{{Spergel}
  et~al.}{2006}]{spergel06}
{Spergel} D.~N.,  et~al., 2006, astro-ph/0603449

\bibitem[\protect\citeauthoryear{{Springel}}{{Springel}}{2005}]{springel05}
{Springel} V.,  2005, \mnras, 364, 1105

\bibitem[\protect\citeauthoryear{{Springel} \& {Hernquist}}{{Springel} \&
  {Hernquist}}{2003}]{springel03}
{Springel} V.,  {Hernquist} L.,  2003, \mnras, 339, 312

\bibitem[\protect\citeauthoryear{{Stark}, {Ellis}, {Richard}, {Kneib}, {Smith}
  \& {Santos}}{{Stark} et~al.}{2007}]{stark07a}
{Stark} D.~P.,  {Ellis} R.~S.,  {Richard} J.,  {Kneib} J.-P.,  {Smith} G.~P.,
   {Santos} M.~R.,  2007, astro-ph/0701279

\bibitem[\protect\citeauthoryear{{Stark}, {Loeb} \& {Ellis}}{{Stark}
  et~al.}{2007}]{stark07b}
{Stark} D.~P.,  {Loeb} A.,    {Ellis} R.~S.,  2007, astro-ph/0701882

\bibitem[\protect\citeauthoryear{{Taniguchi} et~al.,}{{Taniguchi}
  et~al.}{2005}]{taniguchi05}
{Taniguchi} Y.,  et~al., 2005, \pasj, 57, 165

\bibitem[Tapken et al.(2007)]{tapken07} {Tapken} C., {Appenzeller} 
I., {Noll} S., {Richling} S., {Heidt} J., {Meinkoehn} E., \& {Mehlert} D.\ 2007, 
astro-ph/0702414 

\bibitem[Tasitsiomi(2006)]{tasitsiomi06} Tasitsiomi, A.\ 2006, \apj, 
645, 792

\bibitem[\protect\citeauthoryear{{Thommes} \& {Meisenheimer}}{{Thommes} \&
  {Meisenheimer}}{2005}]{thommes05}
{Thommes} E.,  {Meisenheimer} K.,  2005, \aap, 430, 877

\bibitem[\protect\citeauthoryear{{Totani}, {Kawai}, {Kosugi}, {Aoki}, {Yamada},
  {Iye}, {Ohta} \& {Hattori}}{{Totani} et~al.}{2006}]{totani05}
{Totani} T.,  {Kawai} N.,  {Kosugi} G.,  {Aoki} K.,  {Yamada} T.,  {Iye} M.,
  {Ohta} K.,    {Hattori} T.,  2006, \pasj, 58, 485

\bibitem[\protect\citeauthoryear{{White}, {Becker}, {Fan} \& {Strauss}}{{White}
  et~al.}{2003}]{white03}
{White} R.~L.,  {Becker} R.~H.,  {Fan} X.,    {Strauss} M.~A.,  2003, \aj, 126,
  1

\bibitem[\protect\citeauthoryear{{White}}{{White}}{1979}]{white79}
{White} S.~D.~M.,  1979, \mnras, 186, 145

\bibitem[\protect\citeauthoryear{{Willis} \& {Courbin}}{{Willis} \&
  {Courbin}}{2005}]{willis05}
{Willis} J.~P.,  {Courbin} F.,  2005, \mnras, 357, 1348

\bibitem[\protect\citeauthoryear{{Wyithe} \& {Loeb}}{{Wyithe} \&
  {Loeb}}{2006}]{wyithe05b}
{Wyithe} J.~S.~B.,  {Loeb} A.,  2006, \apj, 646, 696

\bibitem[\protect\citeauthoryear{Wyithe, Loeb \& Carilli}{Wyithe
  et~al.}{2005}]{wyithe05}
Wyithe S.,  Loeb A.,    Carilli C.,  2005, \apj, 628, 575

\bibitem[\protect\citeauthoryear{{Zahn}, {Lidz}, {McQuinn}, {Dutta},
  {Hernquist}, {Zaldarriaga} \& {Furlanetto}}{{Zahn} et~al.}{2007}]{zahn06}
{Zahn} O.,  {Lidz} A.,  {McQuinn} M.,  {Dutta} S.,  {Hernquist} L.,
  {Zaldarriaga} M.,    {Furlanetto} S.~R.,  2007, \apj, 654, 12

\bibitem[\protect\citeauthoryear{Zahn, Zaldarriaga, Hernquist \& McQuinn}{Zahn
  et~al.}{2005}]{zahn05a}
Zahn O.,  Zaldarriaga M.,  Hernquist L.,    McQuinn M.,  2005, \apj,
  630, 657

\bibitem[\protect\citeauthoryear{{Zaldarriaga}, {Furlanetto} \&
  {Hernquist}}{{Zaldarriaga} et~al.}{2004}]{zaldarriaga04}
{Zaldarriaga} M.,  {Furlanetto} S.~R.,    {Hernquist} L.,  2004, \apj, 608, 622

\end{thebibliography}

\begin{appendix}

\section{Line Properties}
\label{lineprofile}

Several studies have attempted to understand the impact that
reionization has on the profile of Ly$\alpha$ lines using simple
analytic models \citep{miralda98, santos04, haiman04, furlanetto04c,
dijkstra06b}.  In this section, we investigate the effect of
reionization on the line profile using simulations.  We show that even
when a significant fraction of the universe is neutral, damping wing
absorption owing to neutral patches in the IGM tends to have a small
effect on the average line shape.  We use Method 1 described in
Section \ref{rt} to calculate the intrinsic line profile. 

For simplicity, Method 1 sets the intrinsic width of the line to be
$\Delta \nu = \nu_{\alpha} v_{\rm vir}/c$, where $v_{\rm vir}$ is the
circular velocity of the host halo at the virial radius.  This $\Delta
\nu$ is the width one expects if the typical column density of HI
inside the galaxy, $N_{\rm HI}$, is low enough such that ionizing
photons can escape from the galaxy ($N_{\rm HI} \lesssim 10^{17}
\,{\rm cm}^{-2}$) and, for the massive galaxies of interest, even if
the column density is as high as $N_{\rm HI} = 10^{19-20} \, {\rm
cm}^{-2}$ \citep{dijkstra06b}.  This $\Delta \nu$ is also consistent
with the emergent line profile of a 2-phase ISM in which dense clouds
of HI (with dust inside) scatter Ly$\alpha$ photons.  A 2-phase
structure may be necessary for the transmission of Ly$\alpha$ photons
in the presence of dust \citep{neufeld91, hansen06}.  We do
not attempt to parameterize all of the complicated processes that
determine the width and shape of the intrinsic line, such as the
geometry of the galaxy and its bulk motions.  Such processes can add
additional features to the line profile and make it even more
difficult to disentangle the effect of reionization.

In our calculations, the observed line profiles differ from one
another primarily because of two effects: {\it (1)} The extent along
the line of sight from the emitter to the HII bubble edge. The extent
of the HII region determines the strength of the damping wing
absorption.  This effect is the most important
for our study.  {\it (2)} The amount of infall of gas around the
emitter.  This dictates how much of the red side of the line is
absorbed as some photons on the red side must redshift through
resonance because of the Doppler effect, typically resulting in
absorption.\footnote{The blue side of the line is almost always
entirely absorbed in our calculations.  While most of the blue side of
the line will typically be absorbed in reality, our calculations
overestimate this effect because the simulations do not resolve the
scale on which pressure smooths out density fluctuations.  This lack
of resolution artificially suppresses the transmission.  The
resolution is sufficient to capture the most important effect for this
paper, the damping wing absorption.}  (Outflows -- which we
incorporate with a simple prescription -- have the opposite effect,
allowing more of the blue side of the line to be transmitted.)  The
effect of infall is only crudely modeled here because these
computations are performed on a $0.36$ Mpc grid, whereas the virial
radius of a $5\times10^{10}~\Msun$ halo is $0.09$ Mpc at $z = 7$.
This results in an under-prediction of the effect of infall.\footnote{Analytic
studies typically assume that the infall is maximal at the viral
radius and equal to the circular velocity of the halo \citep{santos04,
dijkstra06b}.  Galactic winds can collisionally ionize
the hydrogen within up to $\sim 0.2$ pMpc of a galaxy, eliminating the
importance of nearby infall \citep{santos04}.}

\begin{figure}
\rotatebox{-90}{\epsfig{file=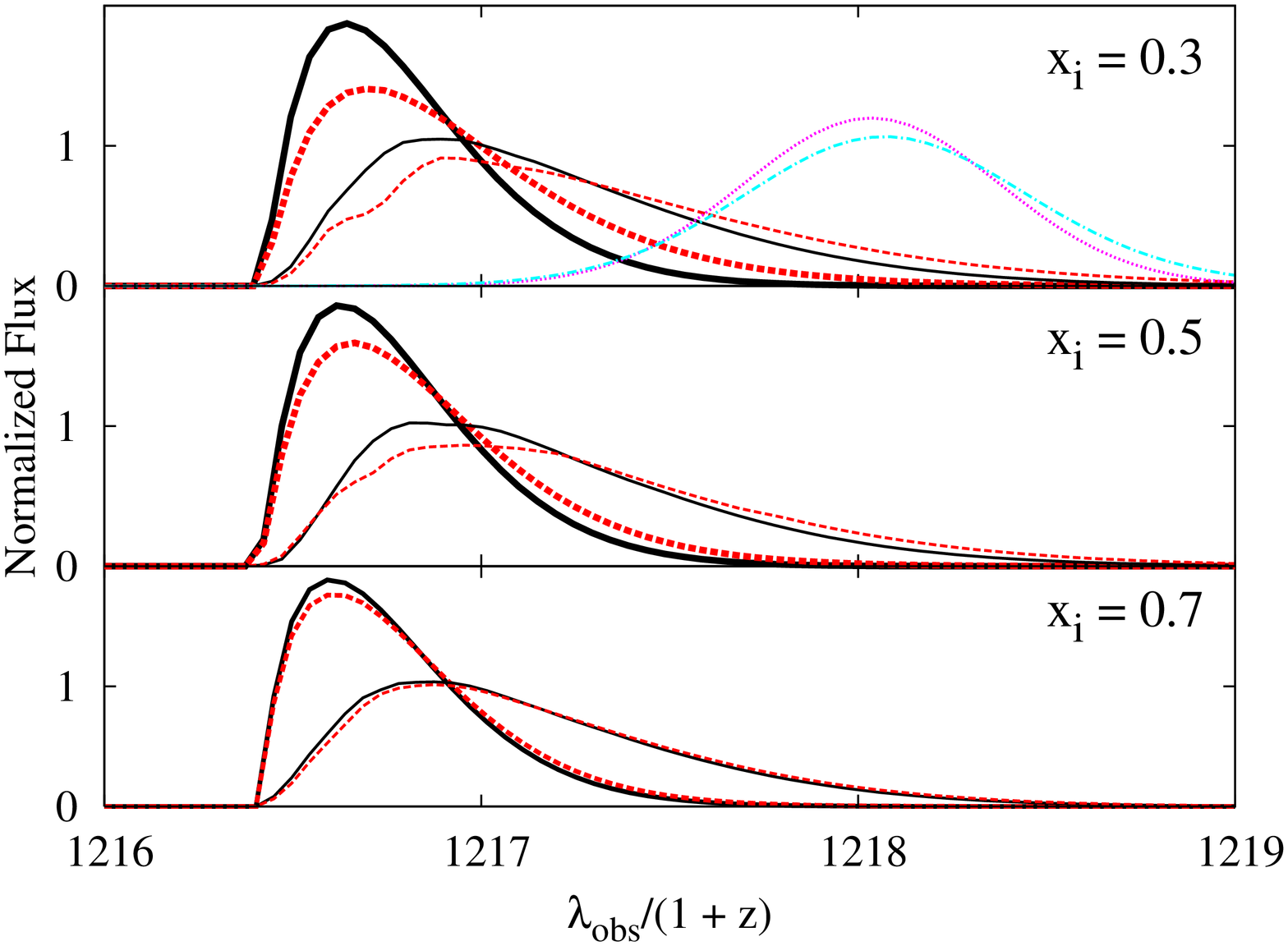, height=9cm}}
\caption{Observed line profile computed from simulation outputs in
which $\barxi = 0.3$ ({\it top panel}), $\barxi = 0.5$ ({\it middle
panel}), and $\barxi = 0.7$ ({\it bottom panel}) for $z = 8.2$, $z =
7.7$, and $z = 7.3$, respectively.  All curves have been normalized
such that their integral is unity to facilitate comparison of the line
profile ({see Fig. \ref{fig:trans} for the relative normalization}).
The thick dashed curves are the average profile of LAEs with $0.5 \,
L_{\rm int, E}(2 \times 10^{10} ~ \Msun) < L_{\rm obs, E} <0.5 \,
L_{\rm int, E}(4 \times 10^{10} ~ \Msun) $ for the quoted $\bar{x}_i$
whereas the thick solid curves are the same for the post-reionization,
ionized-IGM case.  The thin dashed and thin solid curves are the same
as the thick curves but for the LAEs with $0.5 \, L_{\rm int, E}(1
\times 10^{11} ~ \Msun) < L_{\rm obs, E} <0.5 \, L_{\rm int, E}(2
\times 10^{11} ~ \Msun) $.  In the top panel, the dotted and
dot-dashed curves that are offset to the right are respectively the
the $\bar{x}_i \approx 1$ and the $\bar{x}_i = 0.3$ average spectrum
for a toy wind model discussed in the text. These calculations are
performed using the $94$ Mpc simulation of Model (i). }
\label{fig:profiles}
\end{figure}

The quality of the spectrum from a single $z > 6$ LAE for current
ground-based telescopes is never good enough to constrain in detail
the shape of a single Ly$\alpha$ line.  Previous LAE surveys have averaged
the line profile from all of their emitters to generate a higher
quality effective line profile.  Figure \ref{fig:profiles} shows the
average line profiles calculated from snapshots with $\barxi = 0.3$
({\it top panel}), $\barxi = 0.5$ ({\it middle panel}), and $\barxi =
0.7$ ({\it bottom panel}) for $z = 8.2$, $z = 7.7$ and $7.3$,
respectively.  We have normalized the integral over these curves to
unity.  The thick dashed curves are the average profile for LAEs with
$0.5 \,L_{\rm int, E}(2 \times 10^{10} \Msun) < L_{\rm obs, E} < 0.5
\,L_{\rm int, E}(4 \times 10^{10} \Msun)$ for the quoted $\bar{x}_i$,
whereas the thick solid curves are for the same luminosity range but
with $\barxi = 1$.\footnote{The $0.5$ that appears in these bounds is
there to approximate the effect of resonant absorption, which
typically absorbs all frequencies blue-ward of $\nu_0$.}  Because of
the exponential falloff in the halo mass function at large $m$, if we
eliminated the upper bound on $L_{\rm obs, E}$ and included more
luminous LAEs in the average, the line profiles would not be
significantly affected.

Note that the differences between the dashed and solid curves in
Figure \ref{fig:profiles} are small, particularly for the $\bar{x}_i =
0.5$ and $\bar{x}_i = 0.7$ cases.  The reason why reionization does
not have more of an effect on the line profiles is because the average
bubble size in which \emph{observed} LAEs lie is larger than a pMpc
for the $\bar{x}_i$ shown in Figure \ref{fig:profiles}.  Previous
calculations assumed that each LAE was responsible for ionizing its
own bubble, resulting in $R_b < 1 ~$ pMpc (eqn. \ref{eqn:rbauto}) and,
therefore, a larger effect of the damping wing absorption on the
observed line profile (e.g., \citealt{santos04}).  One might wonder
why reionization has any affect on the LAE luminosity function and
correlation function (as seen in the body of this paper) since the
average line profile is not strongly affected.  Even though the line
profile is not changed, the average transmission is decreased by
reionization, and many emitters are not observed because of reduced
transmission.

\begin{figure}
\rotatebox{-90}{\epsfig{file=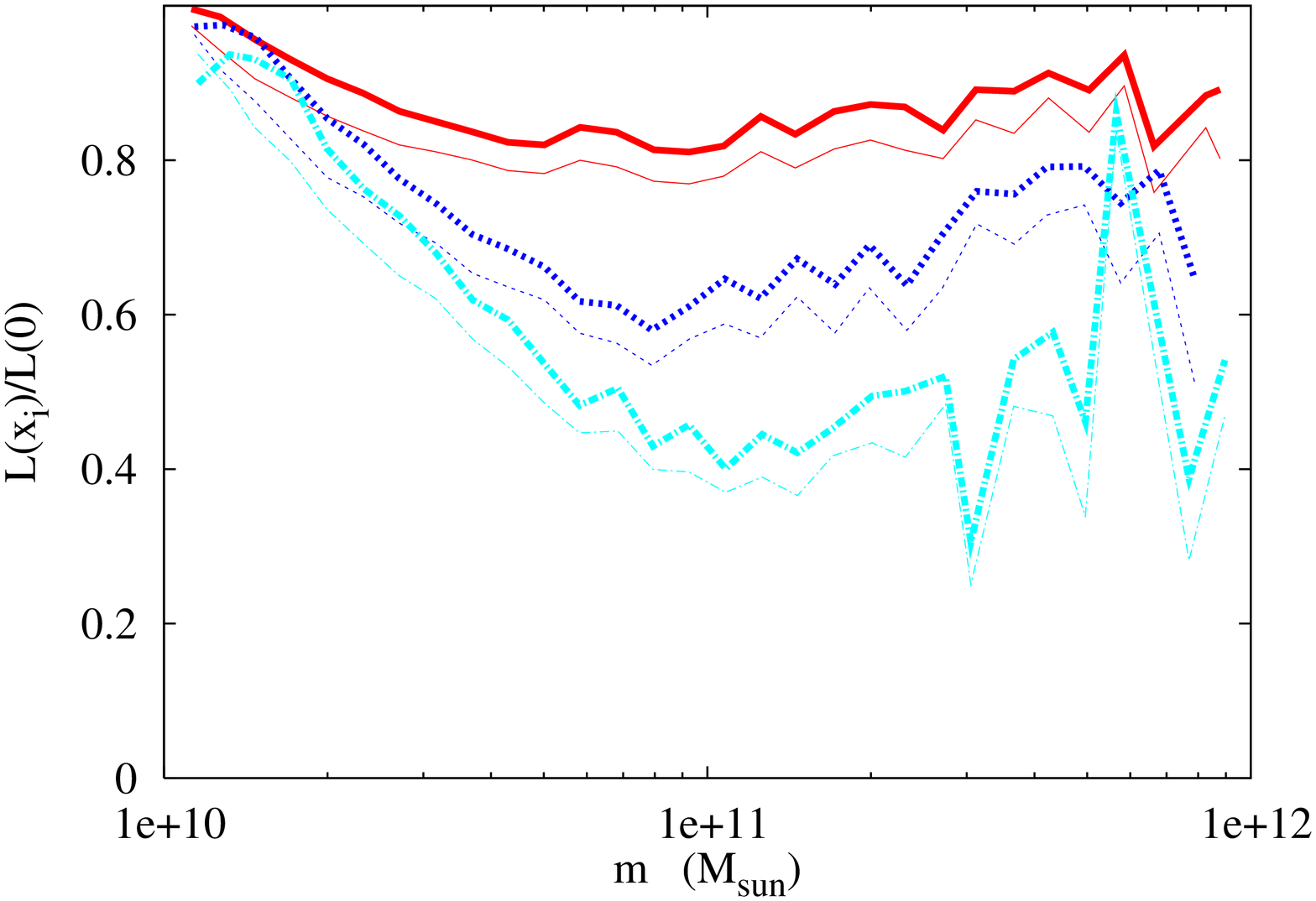, height=8.2cm}}
\caption{Ratio of the average observed luminosity during reionization
to the average observed luminosity for an ionized universe as a
function of halo mass.  The dot-dashed, dotted, and solid thick curves
are computed from simulation snapshots of Model (i) which have
$\bar{x}_i = 0.3$ ($z = 8.2$), $\bar{x}_i = 0.5$ ($z = 7.7$), and
$\bar{x}_i = 0.7$ ($z = 7.3$), respectively, and assuming a survey
that is sensitive to $L_{\rm obs, E} >0.5 \, L_{\rm int, E}(1 \times
10^{10} ~ \Msun)$.  These curves are computed with Method 1 in Section
\ref{calculations}.  The thin curves are the average of
$\exp(-\tau_{\alpha}(\nu_0))$ (as is computed for Method 2 in Section
\ref{calculations}), using the emitters that are observed in the
calculation with Method 1.  The similarity between the thin and thick
curves justifies the use of Method 2.
\label{fig:trans}}
\end{figure}

Galactic winds/outflows, which are found to be very prevalent in lower
redshift LAEs, allow more of the blue side of the line to be
transmitted.  Here we adopt a simple toy model for galactic winds
which is meant to exaggerate their potential effect. We make the same
assumptions for the intrinsic line profile as before, but we redshift
the entire line by the velocity of the wind, which is taken to be
$400$ km s$^{-1}$. This value for $v_{\rm w}$ is motived by the
average velocity offset of the Ly$\alpha$ line in strong emitters at
$z \approx 3$ \citep{shapley03}.  Since galaxies are less massive at
$z >6$ than at $z \approx 3$, the galaxies probably cannot drive such
powerful winds and $v_{\rm w} = 400$ km s$^{-1}$ can be thought of as
an upper bound on $v_{\rm w}$.  Furthermore, it could be the case that
only a fraction of the Ly$\alpha$ photons are scattered by the wind.
The difference between dotted and dot-dashed curves in Figure
\ref{fig:profiles} illustrate the effect of reionization in this wind
model.  Winds reduce the importance of damping wing absorption.

Current narrow band surveys most likely have spectroscopically
confirmed LAEs that have $m \approx 10^{11} ~M_{\odot}$ rather than $m
\approx 2\times 10^{10}~M_{\odot}$, as considered above. (Only the
brightest emitters in a narrow band survey can be spectroscopically
confirmed.) If we take instead emitters within the range $0.5 \,
L_{\rm int, E}(1 \times 10^{11} \; \Msun) < L_{\rm int, E} <0.5 \,
L_{\rm int, E}(2 \times 10^{11} \; \Msun)$ ({\it thin curves in
Fig. \ref{fig:profiles}}), the differences between the line profiles
for $\bar{x}_i \approx 1$ and $\bar{x}_i \approx 0.5$ are even smaller
than in the case previously examined, owing to these more biased
sources sitting preferentially in the largest HII regions.

In our computation, the FWHM of the observed line profiles for LAEs
that have $m \approx 2\times10^{10}~\Msun$ is $\approx 4 \; \AA$ ({\it
thick curves in Fig. \ref{fig:profiles}}) and for those that have $m
\approx 10^{11}~ \Msun$ is $\approx 8 \; \AA$ ({\it thin curves}).
The FWHM of the average line of the spectroscopically confirmed $z=
6.6$ emitters is $\approx 10 \; \AA$, slightly larger than the FWHM in
our calculations \citep{kashikawa06}.  Interestingly,
\citet{kashikawa06} finds a weak anti-correlation between the FWHM of
the line and $L_{\rm obs, E}$.  In our simple model for the LAEs, in
which $\Delta \nu \sim m^{1/3}$, there should be a
correlation. \citet{haiman04} suggest that an anti-correlation may be
a signature of reionization.  However, to reach this result,
\citet{haiman04} assume that the intrinsic widths of the Ly$\alpha$
lines are the same for emitters of all $L_{\rm int, E}$ and that the
HII region around an emitter is created just by this emitter.  The
anti-correlation result from \citet{haiman04} depends on these dubious
assumptions.  There is probably too much uncertain astrophysics to
understand reionization through the correlation of the FWHM with
$L_{\rm obs, E}$. The weak anti-correlation between FWHM with $L_{\rm
obs, E}$ that is observed (with low statistical significance) may
indicates that there is more dispersion in the mapping between $m$ and
$L_{\rm obs, E}$ than we have assumed in this work or that winds are
important in some $z=6.6$ emitters.  As discussed in previous sections,
increasing the dispersion in the mapping between $m$ and $L_{\rm obs,
E}$ does not significantly affect our conclusions pertaining to the
LAE luminosity function and to LAE clustering.

The emitters in a survey whose line profiles are most strongly affected
by reionization may constitute a small sub-sample of the LAEs in a
survey.  Therefore, it might be more fruitful to look at individual
lines rather than the average line profile to detect reionization.
However, even if a survey has sufficient signal to noise to study the
Ly$\alpha$ line from a single emitter, it will be challenging to
isolate the impact of damping wing absorption from other effects.

Thus far, we have considered the effect of a single model for
reionization, model (i), on the line profile. In model (iii), since
the bubbles are smaller, the average line profile at fixed $\bar{x}_i$ is
slightly more altered by reionization than in model (i).  The opposite
is true in model (ii), in which the bubbles are larger.  However, the
differences in the line profiles between the models at fixed
$\bar{x}_i$ are comparable to or smaller than the differences between
the line profiles at $\bar{x}_i = 0.3, 0.5$, and $0.7$ in model (i).

Figure \ref{fig:profiles} ignored the effect of reionization on the
transmission of the Ly$\alpha$ line and just illustrated its effect on
the line profile.  The thick curves in Figure \ref{fig:trans}
represent the ratio of the average of $L_{\rm obs, E}$ at $\bar{x}_i$
to the same but at $\bar{x}_i \approx 1$.  Only emitters with $L_{\rm
obs, E} >0.5 \, L_{\rm int, E}(1 \times 10^{10} ~ \Msun)$ are included
in these averages.  This ratio yields approximately the factor
$\exp(-\tau_{\alpha}(\nu_0))$ -- the strength of damping wing
absorption.  The thick dot-dashed curve is for $\bar{x}_i = 0.3$, the
thick dotted curve is for $\bar{x}_i = 0.5$, and the thick solid curve
is for $\bar{x}_i = 0.7$.  As $\bar{x}_i$ decreases, the average
$L_{\rm obs, E}$ also decreases owing to smaller bubbles and increased
damping wing absorption.  At low $m$, the decrease in $L_{\rm obs,
E}$ with $\bar{x}_i$ is smaller than at large $m$.  This result
might may appear surprising since more massive halos sit in larger
bubbles, which would result in the opposite tendency.  However, this
trend is simply a selection effect owing to the luminosity threshold
of this mock survey.  In addition, this selection effect illustrates
why the line profiles are not affected as substantially as one might
expect: Owing to the nature of the luminosity function, most emitters
sit near the luminosity threshold.  In order for these emitters to be
observed, they must sit in very large HII regions such that the effect
of damping wing absorption is small.

\section{Void Statistics}
\label{voids}

\begin{figure}
\rotatebox{-90}{\epsfig{file=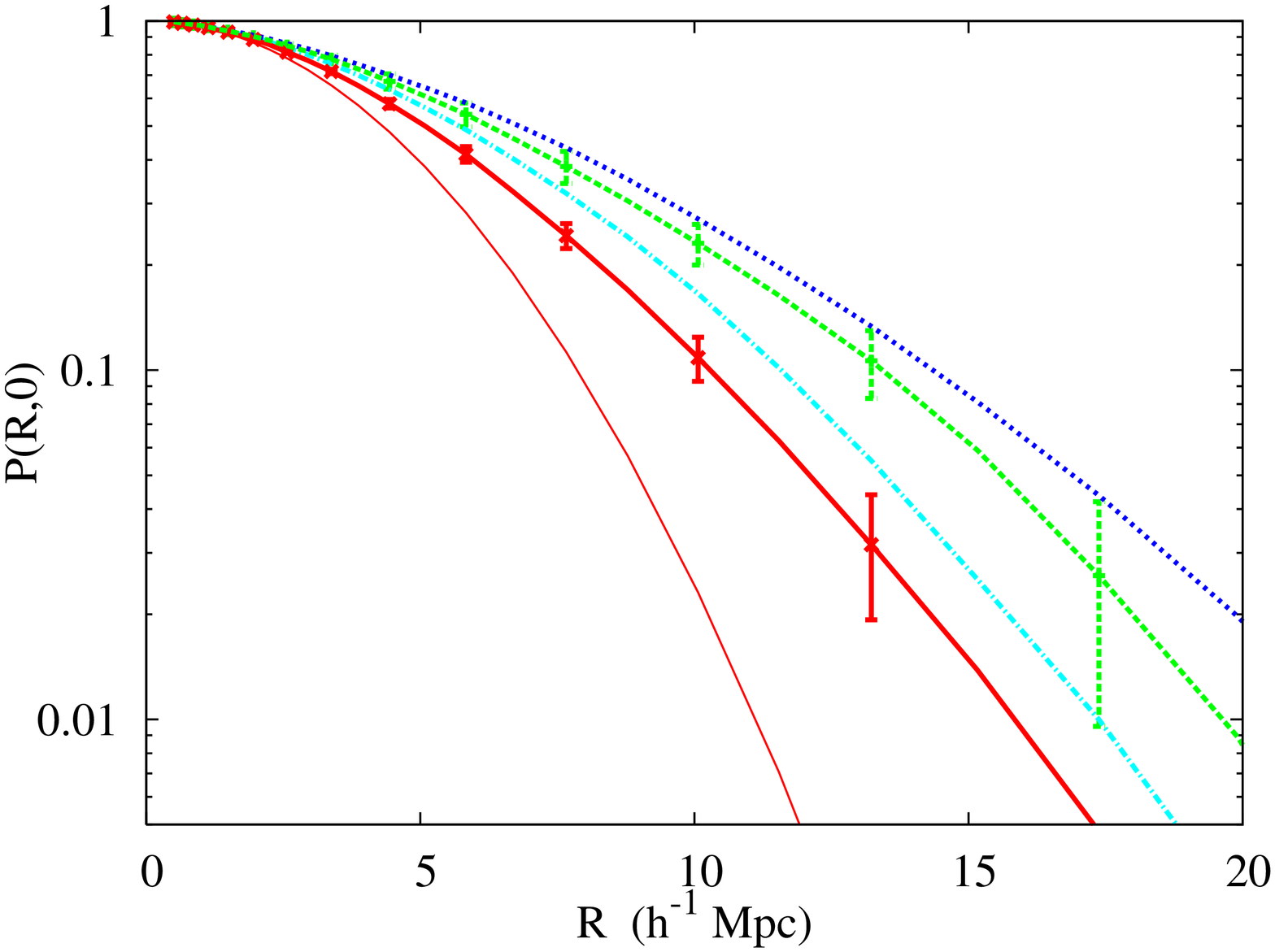, height=8.6cm}}
\caption{2-D void probability distribution function $P(R, 0)$ at
$z= 7.5$ computed from a simulation of model (i). Each curve is
calculated by averaging five independent surveys of $200$ emitters
in a volume of $186 \, \Mpc \times 186 \, \Mpc \times 130 \AA$.  The
error bars are the standard deviation on the curves, calculated from
five mock surveys in independent volumes within the simulation box.
The thin solid curve is $P(R, 0)$ for a purely Poisson
distribution. The thick solid and dot-dashed curves are $P(R, 0)$ for
$\bar{x}_i \approx 1$ with LAEs in halos with $m > 5\times 10^{10} \,
M_{\odot}$ ($f_{\rm E} = 0.15$) and with $m > 1\times10^{11} \, M_{\odot}$
($f_{\rm E} = 0.6$), respectively. The long dashed [short dashed]
curves are the ionized case for $\bar{x}_i = 0.5$ and $m > 3\times
10^{10}~\Msun$ ($f_{\rm E} = 0.15$) [for $\bar{x}_i = 0.5$ and $m >
5\times 10^{10}~\Msun$ ($f_{\rm E} = 0.5$)]. 
\label{fig:vpdf}}
\end{figure}

\begin{figure}
\rotatebox{-90}{\epsfig{file=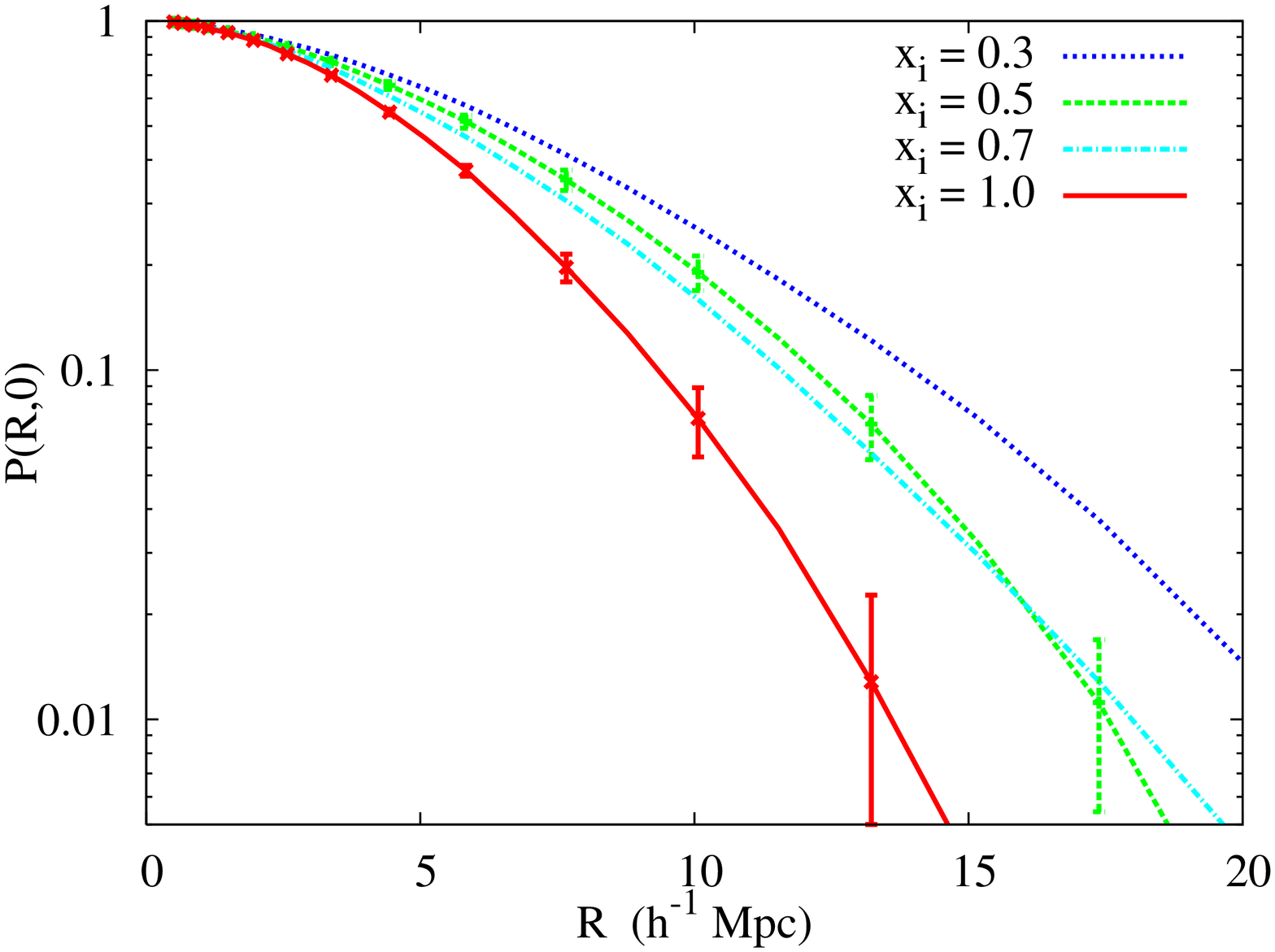, height=8.6cm}}
\caption{The same as \ref{fig:vpdf} but the curves are calculated at
$z = 6.6$ and $m_{\rm min} = 5\times 10^{10}~\Msun$.  This figure
compares the $P(R, 0)$ in reionization model (i) for different
$\bar{x}_i$.
\label{fig:vpdf_xi}}
\end{figure}

The power spectrum may not be the optimal statistic to pick out the
modulation of LAEs owing to the bubbles.  One distinctive feature of
the LAE field during reionization is that there are large voids owing
to the large neutral regions ({see Figs. \ref{fig:maps} and
\ref{fig:maps2}}). This feature motivates using the void probability
distribution function $P(R, 0)$ to measure the effect of reionization,
as was done in \citet{kashikawa06}. Void statistics have a long history
of being used to interpret galaxy surveys (e.g., \citealt{white79}
and \citealt{croton04}). 

$P(R, 0)$ is the probability that a circle of
radius $R$ around a point in the survey field does not contain any
galaxies.  For a Poisson distribution of galaxies, $P(R, 0) =
\exp(-\pi R^2 \,\bar{\Sigma}_{\rm E})$, where $\bar{\Sigma}_{\rm E}$
is the surface density of LAEs.  A generalization of this statistic is
the probability that a region contains $N$ galaxies, $P(R, N)$, and,
for a Poisson distribution, $P(R, N) = (\pi R^2 \, \bar{\Sigma}_{\rm
E})^{N} \, \exp(-\pi R^2 \, \bar{\Sigma}_{\rm E})/N!$.

Figure \ref{fig:vpdf} displays $P(R, 0)$ computed from the $186$ Mpc
simulation of model (i).  Each curve is calculated by averaging five
independent surveys of $200$ emitters in a volume of $186 \, \Mpc
\times 186 \, \Mpc \times 130 \AA$ (a little larger than the volume of the SXDS),
yielding $\bar{n}_{\rm E} = 1.5\times10^{-4} \, \Mpc^{-3}$.  The thin
solid curve is $P(R, 0)$ for a Poisson distribution.  The thick solid
and dot-dashed curves are $P(R, 0)$ of an ionized universe with LAEs
in halos with $m > 5\times 10^{10} \, M_{\odot}$ and with $m >
1\times10^{11} \, M_{\odot}$, respectively. The dashed (dotted curves)
are the ionized case with $\bar{x}_i = 0.5$ and with $m > 3\times
10^{10}~ \Msun$ ($m > 5\times 10^{10} ~\Msun$).  We adjust the duty
cycle to fix the number of emitters, requiring $f_{\rm E} = 0.5$ for
the dotted curve, $f_{\rm E} = 0.15$ for the dashed curve, $f_{\rm E}
= 0.6$ for the dot-dashed, and $f_{\rm E} = 0.15$ for the thick solid.

The error bars in Figure \ref{fig:vpdf} are the standard deviation
from a sample of five independent mock surveys with the specifications
given above.  Note that the error bars in different bins are
correlated.  From the errors, we see that we can distinguish the
models in this plot at $> 1$-$\sigma$ confidence level, and, in
particular, we can distinguish the case with reionization from those
without reionization.  This is promising because the dot-dashed curve
assumes that the LAEs are in the rarest, most clustered halos, and
this curve is below those with $\bar{x}_i = 0.5$.  Clustering of LAEs
owing to reionization generates large-scale voids to a much larger
degree than the intrinsic clustering of halos in $\Lambda$CDM.

Figure \ref{fig:vpdf_xi} shows $P(R, 0)$ for different $\bar{x}_i$
calculated from reionization model (i).  These curves are calculated
assuming $200$ LAE, $m_{\rm min} = 5\times 10^{10}~\Msun$, and
for a survey comparable in volume to the SXDS.  As with the correlation
function, $P(R, 0)$ can distinguish the four different $\bar{x}_i$ in
this figure.  Qualitatively, the significance with which $P(R, 0)$
enables one to distinguish between these different $\bar{x}_i$ appears
to be comparable to the significance the correlation function affords.

Thus far, we have assumed no contamination from foreground galaxies,
which is certainly not the case for photometric LAE surveys.
Unfortunately, a luminosity-limited, widefield spectroscopic LAE
survey at $z > 6$ has not been conducted. If there is significant
contamination in the survey, then the measured value of $P(R,0)$ will
be significantly biased, suppressed (on average) by the factor $\exp(-\pi R^2
\,\bar{\Sigma}_{\rm cont})$ where $\bar{\Sigma}_{\rm cont}$ is the
density of contaminating galaxies (assuming that the contaminants are
uncorrelated). If $\bar{\Sigma}_{\rm cont}/\bar{\Sigma}$ is
appreciable, this suppression is significant, especially at large $R$.
\citet{kashikawa06} measured $P(R,0)$ from the SDF
photometric sample of LAE at $z = 6.6$.  They found that the measured
$P(R,0)$ is consistent with a random sample.  This conclusion may owe
to contamination biasing their measurement. 

We can attempt to alleviate this issue of foreground galaxy
contamination biasing $P(R,0)$ by instead computing the statistics
$P(R, 1),\; P(R, 2)$, etc., which are progressively less biased by
this effect.  In this vein, we can do the exact opposite and use peaks
rather than voids to probe reionization.  As long as these peaks are
sufficiently rare, they will exist only in a distribution of emitters
that is clustered.  In addition, peaks in the emitter distribution are
a distinctive feature of Ly$\alpha$ maps during reionization ({see
Figs. \ref{fig:maps} and \ref{fig:maps2}}).

We have investigated the statistic
\begin{equation}
 P_p(R, \gamma) = \sum_{n ={\rm floor}( \gamma \, \pi R^2
 \,\bar{\Sigma})}^\infty \, P(R, n),
\end{equation}
where $\gamma$ is a constant that we have varied between $3$ and $10$ and
${\rm floor}(x)$ is a function that returns the largest integer
smaller than $x$.  We find that $P_p(R, \gamma)$ can distinguish the
models considered in Figure \ref{fig:vpdf} with comparable
significance to $P(R,0)$.  Like $P(R,0)$, a survey needs an estimate
for its contamination fraction to be able to predict $P_p(R, \gamma)$.
This can be estimated by performing spectroscopic follow-up on a
portion of the survey.  Unlike $P(R,0)$, the bias in the estimation of
$P_p(R, \gamma)$ incurred by foreground contamination is not an
exponential factor, and, therefore, a survey can more accurately
correct for this bias with an estimate of $F_c$.

More work needs to be done to quantify how much independent
information derives from void (and peak) statistics compared to the
correlation function and to quantify the benefits of these statistics
over the correlation function.

\section{Duty Cycle}
\label{duty}

In \citet{mcquinn06b}, the effect of the morphology of reionization on
the source properties, on the redshift of reionization, and on the
minihalos were investigated in detail. \citet{mcquinn06b} found that
the structure of the HII regions during reionization was robust to
most considered effects, with the most important dependence being the
value of the ionized fraction.  However, the effect of a duty cycle
for the ionizing sources -- the fraction of time the sources are
emitting ionizing photons -- on the structure of the HII regions was
omitted in their analysis.  It is probable that star formation at high
redshifts is sporadic.  At any given time, a small fraction of
galaxies may contain the high mass stars that produce ionizing
photons.  These active galaxies are probably also the LAEs.

\begin{figure}
\epsfig{file=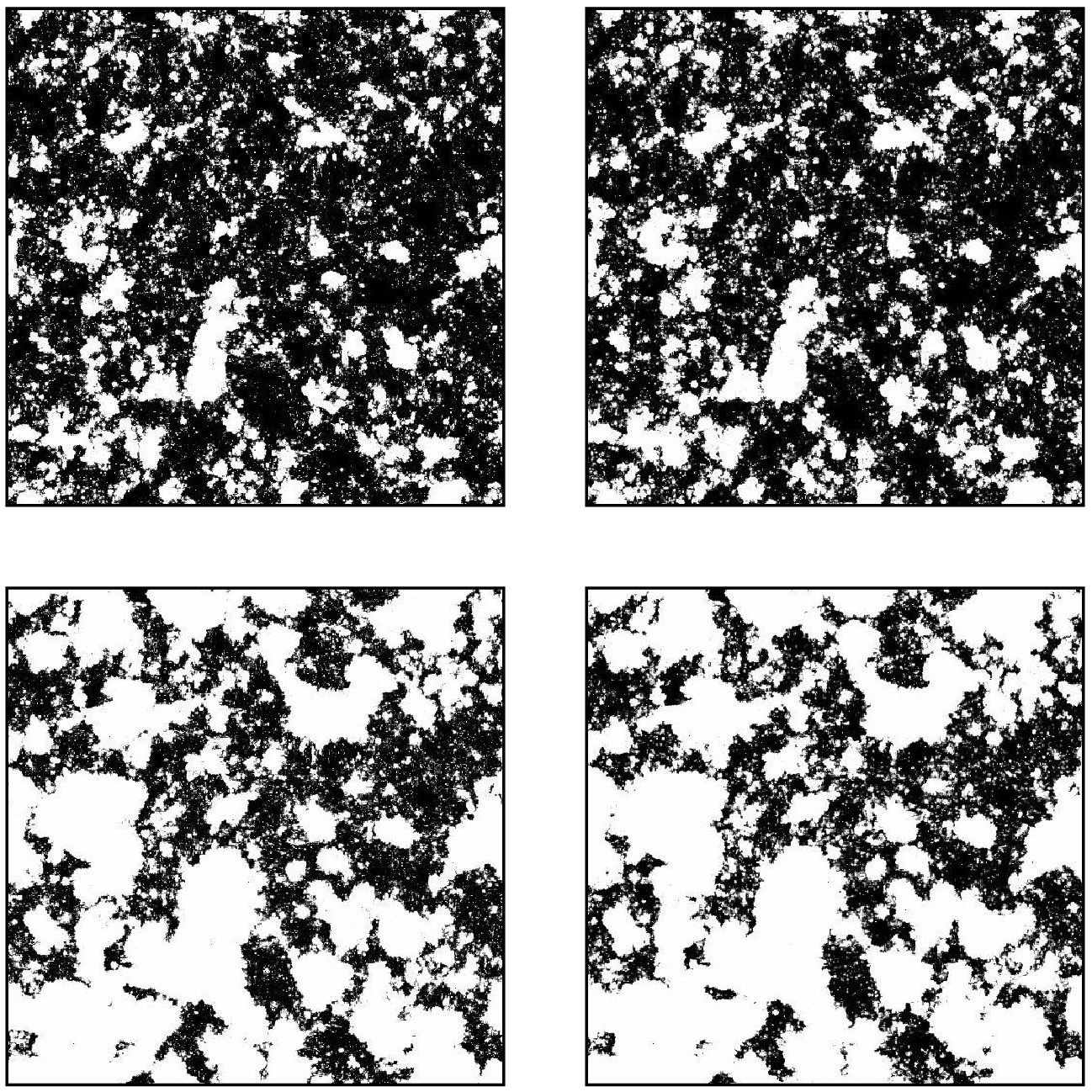, width=8.5cm}
\caption{Effect of the duty cycle of ionizing sources on the
morphology of reionization.  The white regions are ionized and the
black are neutral.  The left panels are from a simulation using model
(i), and the right panels are from an identical simulation except that
only $10\%$ of the sources are active at any time, and these sources
have $10$ times the luminosity compared to the same sources in model
(i).  The active sources are chosen randomly, and this randomization
is performed every $20$ million years.  The top panels have $\bar{x}_i
= 0.3$, and the bottom panels have $\bar{x}_i = 0.7$.  All panels are
from a slice through the $186$ Mpc box. \label{fig:duty}}
\end{figure}

It is interesting to understand the effect of a small ionizing photon
duty cycle $f_{\rm ion} \ll 1$ on the structure of the HII regions.
None of the simulations used in the body of this paper included this
effect.  Figure \ref{fig:duty} compares the $130 ~\Mpc/h$ simulation
using model (i) in which $f_{\rm ion} = 1$ ({\it left panels}) to a
simulation where $f_{\rm ion} = 0.1$ and in which the sources that are
active have a luminosity that is boosted by a factor of $10$ over the
sources in model (i) ({\it right panels}).  To achieve $f_{\rm ion} =
0.1$, $10\%$ of the sources are randomly selected to be active, and
this randomization is performed every $20$ million years.
Twenty-million years was chosen to roughly match the lifetime of
massive stars.  One can see in Figure \ref{fig:duty} that the large
HII regions are very similar between these two cases.  This invariance
owes to thousands of emitters in each large bubble contributing to its
ionization, such that the total number of photons produced inside the
bubble, and, therefore, the bubble size is largely unaffected by
$f_{\rm ion}$.  A similar conclusion about the effect of $f_{\rm ion}$
was reached in the analytic study of \citet{furl-models}.

The duty cycle does influence the smallest HII regions -- bubbles
small enough such that shot noise in the number of galaxies is
important.  The criterion for $f_{\rm ion}$ to change the LAE
statistics from models in which all galaxies are active is that the
fluctuations owing to shot noise in the total number of source
galaxies must be comparable to the cosmological fluctuations in the
number of these galaxies within bubbles of size $\gtrsim 1$ pMpc.  At
$z \sim 7$ there are $\sim 10^4 \, f_{\rm ion}$ galaxies emitting
ionizing photons in an HII region of size $1$ pMpc in the source
models discussed in this paper, whereas cosmological fluctuations in
the source abundance are of the order of unity at this scale.
Therefore, it would require either $f_{\rm ion} \sim 10^{-3}$ for the
value of $f_{\rm ion}$ to affect the statistics of the LAEs.

\section{Source Grouping}
\label{code}
The \citet{sokasian01, sokasian03} and \citet{mcquinn06b} radiative
transfer code has been improved for this paper to group sources more
efficiently, speeding up the computation.  In previous versions of the
code, only sources that fall within the same grid cell were grouped as
a single source, often resulting in tens of millions of sources that
must be processed in a time step.  This algorithm did not take
advantage of the fact that the HII regions become much larger than the
size of a grid cell, motivating a more aggressive grouping
algorithm without a significant loss in accuracy.

Our algorithm for grouping sources is as follows:
\begin{enumerate}
     \item Smooth ionization field $x_i$ at scale $R$ with a top hat filter, yielding the field $\tilde{x}_{i, R}.$
     \item Loop over sources in random order.  \subitem If a
     source is in a cell in which $\tilde{x}_{i, R} > \zeta$,
     check that $\tilde{x}_{i, r} > \zeta$ is also satisfied at all $r
     < R$.  If yes, group sources within a sphere of radius $\eta R$
     into a single source, and place the new source at the center of
     luminosity of the grouped sources.
     \item Repeat previous steps for smaller $R$, but only using the
     sources which have not been grouped.  Stop when $\eta
     \,R$ is less than the width of a grid cell.
     \item Take the new grouped source field and repeat all previous
     steps once again.
\end{enumerate}
In this paper, we start smoothing with $R = 50 ~\Mpc$, and set
$\zeta = 0.9$ and $\eta = 0.1$.  As a test of this algorithm, we have
cross correlated the ionization field generated from a simulation that
has source grouping to one without source grouping, but with
the same sources and at fixed $\bar{x}_i$.  We find no appreciable
difference between the two ionization fields, implying that source
grouping does not affect our conclusions.  Source grouping speeds up
the radiative transfer algorithm considerably when $\bar{x}_i \gtrsim
0.5$.

More aggressive grouping does lead to less accuracy in determining
the local photo-ionization rate around sources, which can influence
the transmission properties of the Ly$\alpha$ line. However, we only
use this more aggressive grouping algorithm in combination with Method
2 in Section \ref{calculations} to calculate the line flux, and Method 2
does not depend on the details of the photo-ionization state of the
gas.

\end{appendix}

\end{document}